\documentclass[11pt]{article}
\usepackage{hyperref}
\usepackage{amsmath}
\usepackage{amssymb}
\usepackage{bm}
\usepackage{latexsym}
\usepackage{graphicx}
\usepackage{euscript} 
\usepackage{mathrsfs}
\begin{document}
\title{Curvature and topology dependency of the cosmological spectra}
\author{Ali A. Asgari* \\Amir H. Abbassi\\Jafar Khodagholizadeh
\\\vspace{6pt} Department of Physics, School of Sciences,\\ Tarbiat Modares University, P.O.Box 14155-4838, Tehran, Iran\\}
\maketitle
\begin{abstract}
In this article we investigate dependency of two important cosmological random fields defined on spatial slices of the FLRW universe and their spectra on the geometry and topology of the background universe. Our discussion includes the post-inflationary universe i.e. radiation-dust mixture era. For this purpose, we first extract an explicit equation describing evolution of the comoving curvature perturbation in the FLRW universe with arbitrary spatial sectional curvature. We may realize when $ K\neq0 $, curvature scale may be as significant as the perturbations scales to recognize the behavior of the spectral indices. We also focus on the entropy perturbation in order to extract behavior of the isocurvature spectral index in terms of the curvature index and time. Our analysis shows that the adiabatic and entropy spectral indices of the cosmological perturbations (spectra of curvature and entropy perturbations) in sub-horizon scales could be function of topology. It may be significant because reveals imprints of geometry on the statistical information deduced by observations. Moreover, an accurate analysis makes clear that time-average of isocurvature index in case $K=0 $ is about zero,so that imprint of entropy perturbation in time duration may be negligible. We also consider evolution of the cosmological indices for super-curvature modes in case $ K=-1 $ . In the most results dependency to curvature, initial conditions and scale modes are thoroughly vivid.
\end{abstract}
\section{Introduction}
\label{sec:intro}

In 1980s inflation theory was proposed by A. Guth as a solution for some long-lasting problems in classical cosmology\cite{1}. However, it was revealed very soon that the inflation could explain the origin of the CMB anisotropy and structure formation\cite{2,3}. As it has been cleared, the quantum fluctuations of a single scalar field in the inflationary epoch (inflaton) are yielded to the classical perturbations which are ascertained as the origin of CMB anisotropy and large structures in the universe\cite{4,5}. The classical perturbations in cosmology could be described by gauge-invariant random fields such as \textit{comoving curvature perturbation} $ \mathcal{R} $ which is associated with the primordial scalar spectrum as well as the conformal factor of the spatial section of the universe\cite{5,6,7}. Evolution of $ \mathcal{R} $ at the inflationary era associated with spatially flat FLRW universe may be described by the \textit{Mukhanov-Sasaki equation}  \cite{8,9} which results in a nearly scale-invariant spectrum\cite{2,6}. The Mukhanov-Sasaki equation has been generalized to include the entire history of the universe\cite{10}. It could be significant because elucidates the evolution of the curvature spectrum during the time. Although observations are consistent with a spatially flat FLRW universe as the background geometry, the current bounds on $ \Omega_K $ doesn't deny the possibility that the universe is a slightly curved\cite{11,12,13}. So the Mukhanov-Sasaki equation can be generalized for the inflationary model with positive curvature index which leads into the resolution of running number problem\cite{14}. In this paper we are going to examine adiabatic and entropy spectra in terms of curvature and topology of the spatial section of the background spacetime. For this purpose, we first acquire a curvature-dependent equation for the evolution of $ \mathcal{R} $ satisfactory for different perturbative scales. Additionally, a discussion of the associated spectral index evolution after inflation in terms of the sectional curvature and time has been presented. We also consider evolution of $ \mathcal{R} $ and its Spectrum for the super-curvature perturbations in case $ K=-1 $ .\\
The outline of this paper is as follow. In section 2, we derive an explicit equation that shows dependency of $ \mathcal{R} $ on curvature and rescaled time which is authentic for all history areas of the universe. Then, we find special solutions of derived equation. In section 4, we focus on the universe containing matter and radiation and study numerical solutions of the $ \mathcal{R} $-evolution equation accompanied with the \textit{Kodama-Sasaki equation}. For this goal, two different initial conditions would be considered. Moreover, we inquire into curvature and topology dependency of the entropy perturbation as the time passes. Finally, the behavior of the spectra in super-curvature modes will be discussed.

\section{Generalized Mukhanov-Sasaki equation}
\label{sec2}

The line element of the universe in the comoving quasi-Cartesian coordinates may be written as \cite{7}
\begin{equation}\label{a1}
ds^2=a^2\left\lbrace - \left( 1 + E\right)d\tau ^2+ 2\partial_i F d\tau dx^i+ \left[ \left( 1 + A\right) \tilde{g}_{ij} + \mathcal{H}_{ij} B\right] dx^idx^j\right\rbrace ,
\end{equation}
where $ \mathcal{H}_{ij}=\bigtriangledown_i\bigtriangledown_j $ is the \textit{covariant Hessian operator} and $ \tilde{g}_{ij}=\delta_{ij}+K\frac{x^i x^j}{1-K\mathbf{x}^2} $ in which $ K\left(=0,\pm1 \right)  $ shows curvature index of spatial slices of the universe. Furthermore, $ E $, $ F $, and $ B $ are scalar random fields describe departure from homogeneity and isotropy and are considered as first order perturbations. Such universe can be split as $ \mathbb{R} \times \mathcal{M} $ where $ \mathcal{M} $ is compact if $ K=+1 $ and non-compact if $ K=0 $ or $ -1 $  (Indeed topology of the universe is $  \mathbb{R} \times \mathcal{M} $ ).\\
On the other hand, the energy-momentum of the cosmic fluid may be decomposed as\cite{7} 
\begin{eqnarray}
&&T_{00}= a^2\left[ \bar \rho\left( 1 + E\right)  + \delta\rho\right] ,\\
&&T_{i0} = a^2\left[\bar p \partial_i F-\left(\frac{\bar \rho+\bar p}{a} \right)\partial _i\left(\delta u \right) \right] ,\\
&&T_{ij} = a^2\left[\bar p\left(1+A \right)\tilde{g} _{ij}+\delta p\tilde{g}_{ij}+ \mathcal{H}_{ij} \left(\bar p B +\Pi ^S\right) \right] , \label{a2}
\end{eqnarray}
where $ \delta u $ and $ \Pi^S $ are the scalar velocity potential and scalar anisotropy inertia of the cosmic fluid, respectively. $ \Pi^S $ measures deviation of the cosmic fluid from perfectness. Moreover, $ \rho=\bar\rho+\delta\rho $ and $ p=\bar p+\delta p $ are energy density and pressure of the cosmic fluid, respectively (Bar over every quantity shows its unperturbed value). One may combine the perturbative scalars to construct gauge-invariant quantities\cite{15}
\begin{eqnarray}
&&\Psi = −\frac{A}{2}-\mathcal{H}\sigma ,\\
&&\Phi =\frac{E}{2}+\mathcal{H}\sigma +\sigma' , \\
&&\mathcal{R} =\frac{A}{2}+H\delta u,\\
&&\zeta =\frac{A}{2}-\mathcal{H}\frac{\delta\rho}{\bar \rho'},\\
&&V =\delta u-a\sigma,\\
&&\Delta =\mathcal{H}\frac{\delta\rho}{\bar \rho'}+H\delta u,\\
&&\Gamma=\delta p-{c_s}^2\delta\rho ,
\end{eqnarray}
with $  \sigma=F-\frac{1}{2}B' $ (shear potential of the cosmic fluid) and $ \mathcal{H}=Ha  $ (comoving Hubble parameter). Here the prime symbol stands for derivative with respect to $ \tau $ and $ {c_s}^2=\frac{d\bar p}{d \bar \rho} $ is the adiabatic sound speed in the cosmic fluid. Now let’s focus on $ \mathcal{R} $ (comoving curvature perturbation) which according to perturbative forms of the Friedmann equations and energy-momentum conservation law depends on the other gauge-invariant potentials (see appendix)
\begin{eqnarray}
\mathcal{R}&=&\frac{-2\mathcal{H}^2+\mathcal{H}'-K}{\mathcal{H}^2-\mathcal{H}'+K}\Psi-\frac{\mathcal{H}}{\mathcal{H}^2-\mathcal{H}'+K}\Psi' +\frac{8\pi G\mathcal{H}^2a^2}{\mathcal{H}^2-\mathcal{H}'+K}\Pi^S ,\label{a3}\\
\mathcal{R}'&=&\frac{-\mathcal{H}{c_s}^2}{\mathcal{H}^2-\mathcal{H}'+K}\bigtriangledown^2\Psi-\frac{4\pi G\mathcal{H}a^2}{\mathcal{H}^2-\mathcal{H}'+K}\left( \Gamma +\bigtriangledown^2\Pi^S\right)\nonumber\\
&-&\frac{K}{\mathcal{H}^2-\mathcal{H}'+K}\left[\Psi'+\mathcal{H}\left( 1+3{c_s}^2\right)\Psi  \right] .\label{a4}
\end{eqnarray}
Here $ \bigtriangledown^2 $ is the Laplace-Beltrami operator associated with spatial folia in the universe which is defined as contracted Hessian operator with $ \tilde{g}_{ij} $
\begin{displaymath}
\bigtriangledown^2=\tilde{g}^{ij}\mathcal{H}_{ij} .
\end{displaymath}
By taking these two equations to the Fourier space i.e. using Fourier transformation on $ \mathcal{M} $ (spatial slice of the universe) one can rewrite equations (\ref{a3}) and (\ref{a4}) as\cite{7,16,17}
\begin{eqnarray}
\mathcal{R}_q&=&\frac{-2\mathcal{H}^2+\mathcal{H}'-K}{\mathcal{H}^2-\mathcal{H}'+K}\Psi _q-\frac{\mathcal{H}}{\mathcal{H}^2-\mathcal{H}'+K}\Psi' _q +\frac{8\pi G\mathcal{H}^2a^2}{\mathcal{H}^2-\mathcal{H}'+K}\Pi^S _q ,\label{a5}\\
\mathcal{R}'_q&=&\frac{\mathcal{H}{c_s}^2\left(q^2-K\right)}{\mathcal{H}^2-\mathcal{H}'+K} \Psi _q-\frac{4\pi G\mathcal{H}a^2}{\mathcal{H}^2-\mathcal{H}'+K}\left[  \Gamma _q -\left(q^2-K\right)\Pi^S _q\right] \nonumber\\
&-&\frac{K}{\mathcal{H}^2-\mathcal{H}'+K}\left[\Psi' _q+\mathcal{H}\left( 1+3{c_s}^2\right)\Psi _q \right] .\label{a6}
\end{eqnarray}
Here Fourier transform can be thought as decomposition in terms of eigen functions of the Laplace-Beltrami operator \cite{7,16}. Note the index $ q $ stands for the Fourier transforms and refers to the perturbation scale. Additionally, $ q $ is non-negative and continuous for $ K=0 $ and $ -1 $ while natural (discrete) for case $ K=+1 $ as a direct consequence of compactness of $ \mathcal{M} $. By a tedious calculation it is possible to combine equations (\ref{a5}) and (\ref{a6}) to derive an explicit equation in terms of $ \mathcal{R}_q $
\begin{multline}\label{a7}
\mathcal{R}''_q+2\frac{\mathscr{D}'_q}{\mathscr{D}_q}\mathcal{R}'_q+\left[K\left(\frac{\mathscr{B}'_q}{\mathcal{H}\mathscr{B}_q}-\frac{\mathscr{A}'}{\mathcal{H}\mathscr{A}} \right)+{c_s}^2\left( q^2-4K\right) -K\right] \mathcal{R}_q=\\
\qquad4\pi Ga^2\left[\left( \mathcal{H}\frac{\mathscr{B}'_q}{\mathscr{B}_q}-4\mathcal{H}^2-\mathcal{H}'-K\right)\frac{\mathscr{C}_q}{\mathscr{A}}-\mathcal{H}\frac{\mathscr{C}'_q}{\mathscr{A}}+2\frac{\mathscr{B}_q}{\mathscr{A}}\Pi^S_q  \right],
\end{multline}
where
\begin{eqnarray}
&&\mathscr{A}=\mathcal{H}^2-\mathcal{H}'+K ,\\
&&\mathscr{B}_q={c_s}^2\left(q^2-4K \right)\mathcal{H}^2+K\mathscr{A}, \\
&&\mathscr{C}_q=\Gamma _q-\left(q^2-3K \right)\Pi^S_q ,\\
&&\mathscr{D}_q=a\sqrt{\frac{\mathscr{A}}{\mathscr{B}_q}} .
\end{eqnarray}
Equation(\ref{a7}) is the most general equation which specifies time evolution of $ \mathcal{R}_q $ responsible not only for the inflation duration and spatially flat universe but also for all other epochs in the history of the universe with arbitrary curvature index. Inserting $ K=0 $ in equation (\ref{a7}) vividly results in equation (2.4) in \cite{10} which describes the $ \mathcal{R}_q $–evolution in the spatially flat FLRW universe. We refer equation (\ref{a7}) as the \textit{generalized Mukhanov-Sasaki equation}.

\section{Special solutions}
\label{sec3}

\begin{figure}[!htb]
\centering
\includegraphics[width=.5\textwidth]{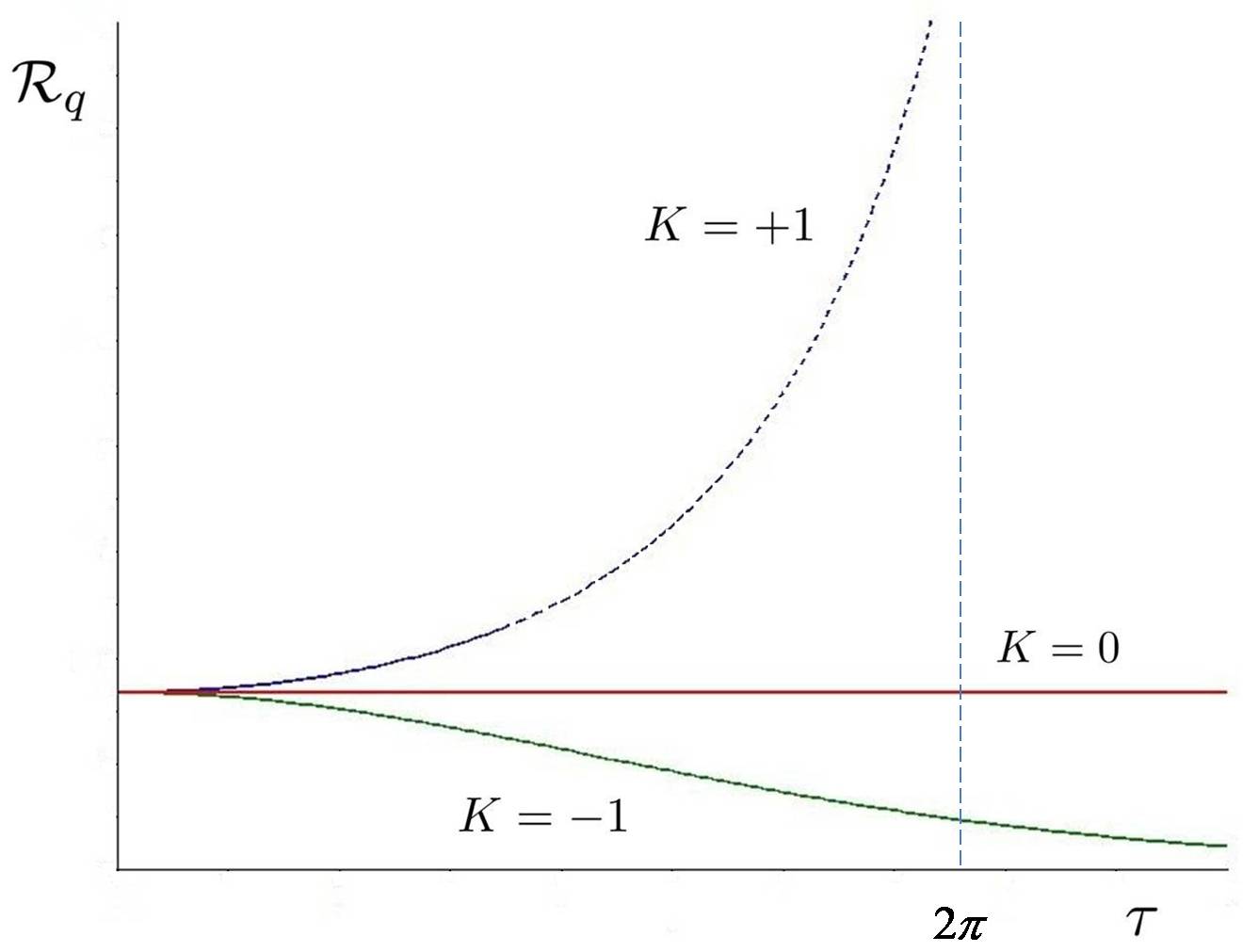}
\caption{Evolution of  $ \mathcal{R}_q $ in a universe constructed from dust for  $ K=0,\pm1 $. Obviously, $ \mathcal{R}_q $ is independent of perturbation scales. For open topology cases ($ K=0 $ and $ K=-1 $) $ \tau $ takes all the values from zero to infinity whereas for closed topology case ($ K=+1 $ ) just values in the interval $ \left( 0,2\pi \right)  $ are permissible for $ \tau $. The asymptote of the cure in case $ K=+1 $ accentuates this point.}\label{fig1D}
\end{figure}

\begin{figure}[!htb]
\centering
\includegraphics[width=1\textwidth]{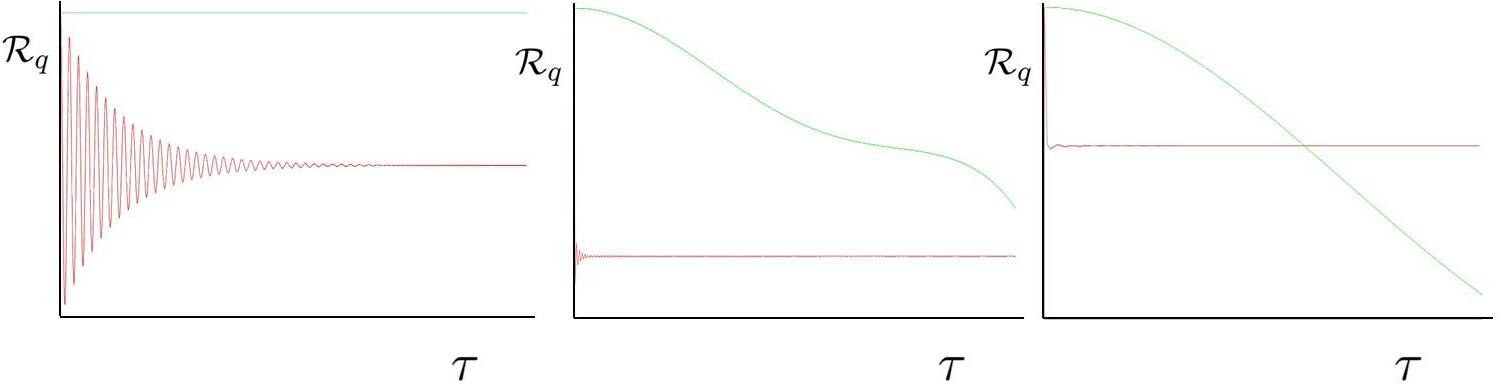}
\caption{Evolution of  $ \mathcal{R}_q $ in a radiation-contained universe for  $ K=0 $ (left), $ K=+1 $ (center) and $ K=-1 $ (right). "Red" curves are related to a typical sub-horizon mode ($ q\gg1 $)  and "green" curves show the behavior of $ \mathcal{R}_q $ at super-horizon modes ($ q\ll 1 $ for $ K=0,-1 $ and $ q\sim 4 $ for $ K=+1 $). Similar to figure 1, for $ K=0$ and $ K=-1 $ cases $ \tau $ can admit all positive values however for $ K=+1 $ case we have $ 0<\tau<\pi $.}\label{fig1R}
\end{figure}

As cited before, there are new observational evidences which indicate that the sectional curvature of the universe is not zero\cite{12,13}. On the other hand, if $ K\neq 0 $ the curvature scale $ \frac{\vert K\vert}{\mathcal{H}^2\left(t \right) }=\vert\Omega_K\left(t \right) \vert $ would be significant like the perturbation mode scales. The curvature scale may be expressed as the sectional curvature of the spatial slices of the universe $ \mathcal{K}_s $ because $ \Omega_K\left(t \right)=-\frac{ \mathcal{K}_s\left(t \right) }{H^2\left( t\right) }  $ . \\
now let’s focus on the universe filled by pure dust, so
\begin{displaymath}
\Pi^S=0\quad,\quad\Gamma=0\quad,\quad{c_s}^2=0 .
\end{displaymath}
Consequently, equation (\ref{a7}) reduces to
\begin{equation}\label{a8}
\mathcal{R}''_q+2\mathcal{H}\mathcal{R}'_q-\mathcal{R}_q=0 ,
\end{equation}
where $ \mathcal{H}=\sqrt{K}\cot\left( \frac{\sqrt{K}}{2}\tau\right)  $ . The solution of equation (\ref{a8}) is  
\begin{equation}\label{a9}
\mathcal{R}_q\propto \frac{-\frac{\tau}{2}\cos\left( \frac{\sqrt{K}}{2}\tau\right)+\frac{1}{\sqrt{K}}\sin\left( \frac{\sqrt{K}}{2}\tau\right)}{\frac{1}{\sqrt{K}}\sin^3\left( \frac{\sqrt{K}}{2}\tau\right)} .
\end{equation}
Note that equation (\ref{a8}) has another solution which is singular at $ \tau=0 $ , so we condone it. The behavior of $ \mathcal{R}_q $ for $ K=0,\pm1 $ is depicted in figure \ref{fig1D}. 
In case $ K=+1 $, $ \mathcal{R}_q $ increases gradually as time passes and apparently diverges because of irregular behavior of $ \mathcal{H}=\cot \frac{\tau}{2} $. So $ \mathcal{R}_q $ does not remain perturbative and henceforth figure 1 is not plausible due to deflecting from linearity procedure (deviation from linearity has illustrated in figure 1 by dashed part of plot). Figure 1 also elucidates the structure formation in the universe with compact spatial section takes place more quickly than other cases because the power spectral function of the structure which gives rise to variance of the fractional density fluctuations relates to $ \mathcal{R}_q $ \cite{6}. Moreover, perturbations deviate from the linearity rapidly because of different topological structure. Obviously $  \mathcal{R}_q $ is independent of perturbation scales in spite of curvature scale, i.e. in epochs in which $ \vert\Omega_K\left(t \right) \vert\ll1 $ equation (\ref{a8}) results in $ \mathcal{R}_q=const $ regardless of $ K $.\\
Conversely, in a radiated-dominated universe
\begin{displaymath}
\Pi^S=0\quad,\quad\Gamma=0\quad,\quad{c_s}^2=\frac{1}{3} .
\end{displaymath}
So equation (\ref{a7}) can be written as
\begin{multline}\label{a10}
\mathcal{R}''_q+ \frac{2\sqrt{K}\left(q^2+2K \right)\cot\sqrt{K}\tau\left(1+\cot^2\sqrt{K}\tau \right)}{6K+\left(q^2+2K \right)\cot^2\sqrt{K}\tau} \mathcal{R}'_q+\\
\left[\frac{q^2-K}{3}-\frac{2K\left(q^2+2K \right)\left( 1+\cot^2\sqrt{K}\tau\right)}{6K+\left(q^2+2K \right)\cot^2\sqrt{K}\tau } \right] \mathcal{R}_q=0.
\end{multline}
Equation (\ref{a10}) cannot be solved analytically. However, it is possible to perceive the behavior of the solution subject to the plausible initial condition (for example, $ \mathcal{R}_q\left(\tau=0 \right) =const $ and $  \mathcal{R}'_q\left(\tau=0 \right) =0 $) via numerical methods (see figure \ref{fig1R}).  As it can be seen the solution depends on the perturbation scales.
Now let’s turn to the inflation era. Inflation may be treated as a perfect fluid for which
\begin{displaymath}
\Pi^S=0\quad,\quad\Gamma=0\quad,\quad{c_s}^2=1 .
\end{displaymath}
So
\begin{equation}\label{a11}
\mathcal{R}''_q+2\frac{\mathscr{D}'_q}{\mathscr{D}_q}\mathcal{R}'_q+\left[K\left(\frac{\mathscr{B}'_q}{\mathcal{H}\mathscr{B}_q}-\frac{\mathscr{A}'}{\mathcal{H}\mathscr{A}} \right)+q^2-5K\right] \mathcal{R}_q=0,
\end{equation}
which is the Mukhanov-Sasaki equation for general $ K $, has been derived in \cite{18} for $ K=+1 $. We leave solutions of equation (\ref{a11}) for next works and now turn to investigate the evolution of $ \mathcal{R}_q $ in post-inflation epoch.    

\section{Evolution of  $ \mathcal{R}_q  $ in a more realistic universe}
\label{sec4}

Here we are going to investigate the evolution of $  \mathcal{R}_q $ in a more concrete model of the universe in which the Cosmic fluid is a mixture of dust and radiation without any interaction. It means there is no energy and momentum transfer between them. This model was used by Seljak\cite{19} to analyze the CMB anisotropy. As a comparison to the real universe neutrons and photons belong in radiation and CDM is a member of dust part. Obviously, $ \bar{\rho}=\bar{\rho}_M+\bar{\rho}_R $ in which $ \bar{\rho}_M\propto a^{-3} $ and $  \bar{\rho}_R\propto a^{-4} $. One may define the normalized scale factor as
\begin{displaymath}
y=\frac{a}{a_{eq}}=\frac{\bar{\rho}_M}{\bar{\rho}_R}.
\end{displaymath}
Here $ a_{eq} $ is scale factor in the time of matter-radiation equality. It is not hard to show for the composite cosmic fluid
\begin{equation}\label{a12}
\omega =\frac{1}{3\left( y+1\right) } \hspace{17pt} , \hspace{17pt} {c_s}^2=\frac{4}{3\left( 3y+4\right) } \hspace{17pt} , \hspace{17pt} \Pi^S=0 .
\end{equation}
Besides, according to the Friedmann equation one can show
\begin{equation}\label{a13}
\mathcal{H} =\frac{y'}{y}=\frac{\sqrt{-2Ky^2+\left(\mathcal{H}_{eq}^2+K \right) \left(y+1 \right) }}{\sqrt{2}y},
\end{equation}
 in which $ \mathcal{H}_{eq} $ is the comoving Hubble parameter of matter-radiation equality. On the other hand, it can be shown that 
\begin{equation}
\Gamma =-\bar{\rho}_M{c_s}^2\mathcal{S} ,
\end{equation}
where $ \mathcal{S}=\delta_M-\delta_R=\frac{\delta\rho_M}{\bar{\rho}_M}-\frac{3}{4}\frac{\delta\rho_R}{\bar{\rho}_R} $ is the entropy perturbation between matter and radiation. Consequently,
\begin{equation}\label{a14}
\Gamma _q =-\frac{\mathcal{H}_{eq}^2+K}{4\pi Ga_{eq}^2}\frac{\mathcal{S}_q}{y^3\left( 3y + 4\right)}.
\end{equation}
 By substituting equations (\ref{a12}), (\ref{a13}), and (\ref{a14}) in equation (\ref{a7}) we find
\begin{multline}\label{a15}
\mathcal{R}^{\star\star}_q+
\bigg\{ -\frac{8\left( q^2-4K\right)\left(-4Ky+\mathcal{H}^2_{eq}+K \right) +18K\left(\mathcal{H}^2_{eq}+K\right) \left(3y+4 \right)}{8\left( q^2-4K\right) \left[-2Ky^2+\left(\mathcal{H}^2_{eq}+K \right)\left(y+1 \right)\right] +3K\left(\mathcal{H}^2_{eq}+K\right) \left(3y+4 \right) ^2}+\\
\frac{\left(\mathcal{H}^2_{eq}+K \right) \left(y+2\right) }{2y\left[2Ky^2-\left(\mathcal{H}^2_{eq}+K \right)\left(y+1 \right)\right]}+\frac{3}{y}+\frac{6}{3y+4}\bigg\}  \mathcal{R}^{\star}_q+
\frac{2}{-2Ky^2+\left(\mathcal{H}^2_{eq}+K \right) \left(y+1 \right) }\bigg\{-K+Ky\times \\
\frac{8\left( q^2-4K\right)\left(-4Ky+\mathcal{H}^2_{eq}+K \right) +18K\left(\mathcal{H}^2_{eq}+K\right) \left(3y+4 \right)}{8\left( q^2-4K\right) \left[-2Ky^2+\left(\mathcal{H}^2_{eq}+K \right)\left(y+1 \right)\right] +3K\left(\mathcal{H}^2_{eq}+K\right) \left(3y+4 \right) ^2}+\frac{4\left(q^2-4K \right) -18Ky}{3\left(3y+4 \right) } \bigg\} \mathcal{R}_q\\
=\frac{4}{\left( 3y+4\right) ^2\left[ -2Ky^2+\left(\mathcal{H}^2_{eq}+K \right) \left(y+1 \right)\right] }\bigg\{-\frac{72Ky^3+80Ky^2-\left(\mathcal{H}^2_{eq}+K\right) \left( 39y^2+80y+40\right) }{2y\left(3y+4\right) }\\
-2\frac{\left[-2Ky^2+\left(\mathcal{H}^2_{eq}+K \right) \left(y+1 \right)  \right] \left[4\left( q^2-4K\right)\left(-4Ky+\mathcal{H}^2_{eq}+K \right)+9K \left(\mathcal{H}^2_{eq}+K\right) \left(3y+4 \right)\right] }{8\left( q^2-4K\right) \left[-2Ky^2+\left(\mathcal{H}^2_{eq}+K \right) \left(y+1 \right)\right] +3K \left(\mathcal{H}^2_{eq}+K\right) \left(3y+4 \right)^2}\bigg\}\mathcal{S}_q\\
-48\frac{y+1}{y\left(3y+4 \right) ^3}\mathcal{S}_q+\frac{4}{\left(3y+4 \right) ^2}\mathcal{S}^{\star}_q
\end{multline}
Here ''$ \star $'' stands for the partial derivative with respect to $ y $. It is clear from equation (\ref{a15}) that $ \mathcal{R}_q $-evolution depends on $ \mathcal{S}_q $-evolution directly. On the other hand, $ \mathcal{S}_q $ in a universe contained radiation and dust obey from the Kodama-Sasaki equation\cite{20}
\begin{equation}\label{a16}
\mathcal{S}''_q+3{c_s}^2\mathcal{H}\mathcal{S}'_q-\frac{1}{3}\left(3{c_s}^2-1 \right) \left( q^2-K\right) \mathcal{S}_q =-\left( q^2-K \right)\Delta_q , 
\end{equation}
where $ \Delta_q  $ depends on the Bardeen's potential via the Poisson's equation
\begin{equation}\label{a17}
\Delta_q=\frac{q^2-4K}{3\left(\mathcal{H}^2-\mathcal{H}'+K \right)}\Psi_q .
\end{equation}
By combination of equations (\ref{a5}), (\ref{a6}), and (\ref{a17}) we have
\begin{multline}\label{a18}
\Delta_q=\frac{q^2-4K}{3\left[\mathcal{H}^2{c_s}^2\left(q^2-4K \right)+K\left(\mathcal{H}^2-\mathcal{H}'+K\right) \right] }\Big\{ \mathcal{H}\mathcal{R}'_q-\mathcal{R}_q+\frac{4\pi G\mathcal{H}^2a^2}{\mathcal{H}^2-\mathcal{H}'+K}\big[\Gamma_q-\left(q^2 -K\right) \Pi^s_q \big] \Big\} .
\end{multline}
Now by inserting equation (\ref{a18}) into equation (\ref{a16}) and rewriting equation (\ref{a16}) in terms of $ y $ we find
\begin{multline}\label{a19}
\mathcal{S}^{\star\star}_q+\frac{1}{y}\bigg\{ \frac{\left(\mathcal{H}^2_{eq}+K \right) \left(y+2 \right)}{4Ky^2-2\left(\mathcal{H}^2_{eq}+K \right) \left(y+1 \right)}+\frac{4}{3y+4}+1 \bigg\}\mathcal{S}^{\star}_q=\frac{2\left(q^2-K \right) }{2Ky^2-\left(\mathcal{H}^2_{eq}+K \right) \left(y+1 \right)}  \times \\
\bigg\{\frac{y}{3y+4}\mathcal{S}_q+\frac{2y\left(3y+4 \right) \left(q^2-4K \right) }{8\left(q^2-4K \right) \left[-2Ky^2+\left(\mathcal{H}^2_{eq}+K \right) \left(y+1 \right)\right] +3K\left(\mathcal{H}^2_{eq}+K\right) \left(3y+4 \right)^2 }\times\\
\bigg[ \left(-2Ky^2+\left(\mathcal{H}^2_{eq}+K \right) \left(y+1 \right) \right) \left(\mathcal{R}^{\star}_q- \frac{4}{\left(3y+4 \right) ^2}\mathcal{S}_q \right) -2y\mathcal{R}_q\bigg]   \bigg\} .   
\end{multline}
Indeed, equations (\ref{a15}) and (\ref{a19}) are coupled and must be solved simultaneously. Besides, in early stage (more accurately at the end of inflation) remarkable part of perturbations are placed outside the horizon and the role of curvature may be insignificant\cite{21} i.e. 
\begin{equation}\label{a20}
\frac{\vert K\vert}{\mathcal{H}^2}\ll 1 \quad \rm{and}\quad \frac{q}{\mathcal{H}}\ll1 .
\end{equation}
Under these conditions in question system has two outstanding solutions
\begin{itemize}
\item {\bf Solution 1}
\begin{eqnarray*}
\left\{\begin{aligned}
&\mathcal{S}_q=0,\\ \\
&\mathcal{R}_q=const.
\end{aligned}\right.
\end{eqnarray*}
\item {\bf Solution 2}
\begin{eqnarray*}
\left\{\begin{aligned}
&\mathcal{S}_q=const,\\ \\
&\mathcal{R}_q=\frac{y}{3y+4}\mathcal{S}_q=\frac{1}{3}\left( 1-3{c_s}^2\right) \mathcal{S}_q.
\end{aligned}\right.
\end{eqnarray*}
\end{itemize}
These solutions are plausible around $ y=0 $ so they specify the initial conditions for the system. Strictly speaking, the first solution which according to the inflationary theory may be written as\cite{6}
\begin{eqnarray}\label{a21}
y\sim 0:\left\{\begin{aligned}
&\mathcal{S}_q\sim 0 ,\\  \\
&\mathcal{R}_q\sim Nq^{-2+\frac{n_{s0}}{2}} \quad \left(N\simeq 10^{-5}\quad \rm{and}\quad n_{s0}\simeq 0.96 \right)
\end{aligned}\right.
\end{eqnarray}
is referred as the \textit{adiabatic initial condition}. On the contrary, solution 2 is called the \textit{isocurvature initial condition} in accordance with the \textit{Liddle-Mazumdar model}\cite{22} may be written as
\begin{eqnarray}\label{a22}
y\sim 0:\left\{\begin{aligned}
&\mathcal{S}_q\sim Mq^{-2+\frac{n_{iso0}}{2}} ,\quad \left(M\simeq 10^{-5}\quad \rm{and}\quad n_{iso0}\simeq 4.43\right) \\ \\ 
&\mathcal{R}_q\sim \frac{y}{3y+4}\mathcal{S}_q  ,
\end{aligned}\right.
\end{eqnarray}

\begin{figure}[!htb]
\centering
\includegraphics[width=.49\textwidth]{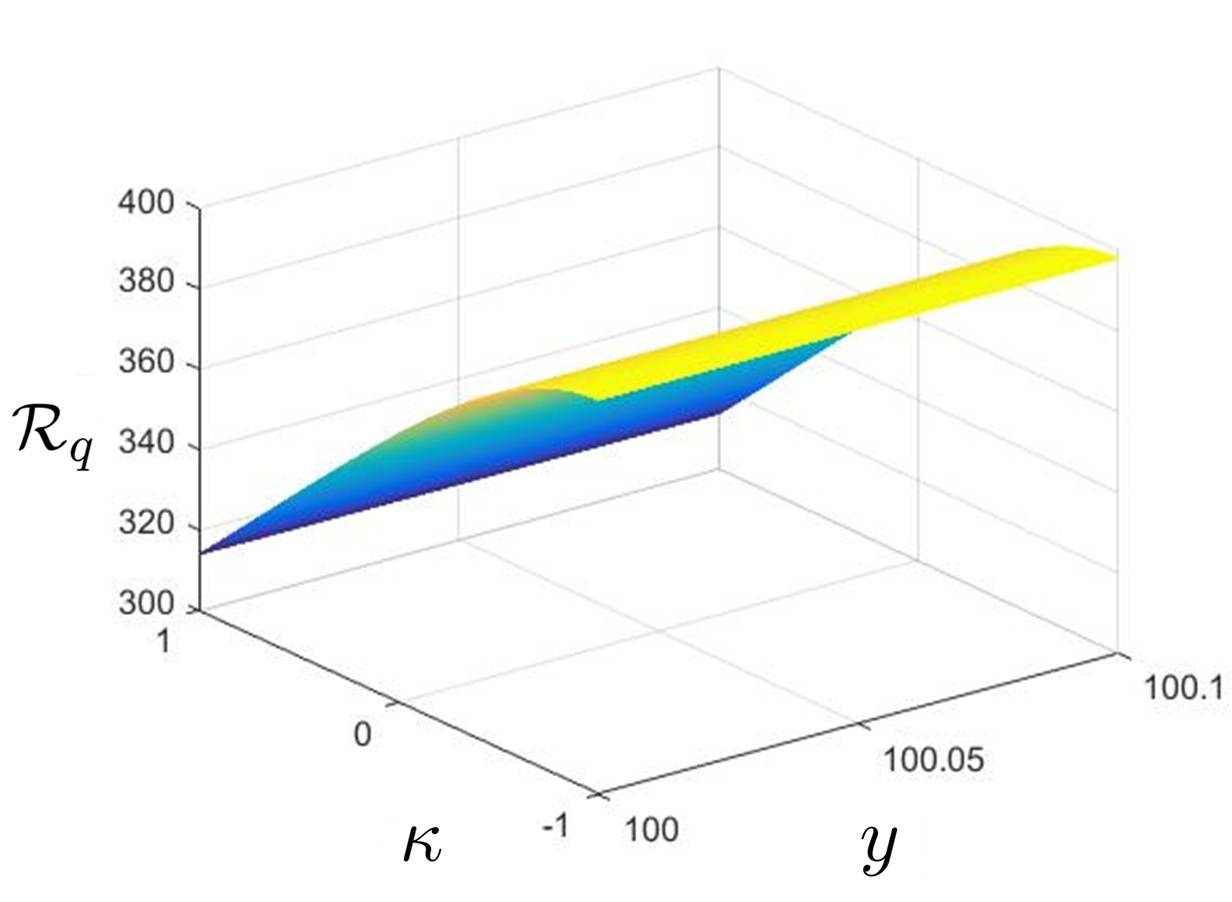}
\includegraphics[width=.49\textwidth]{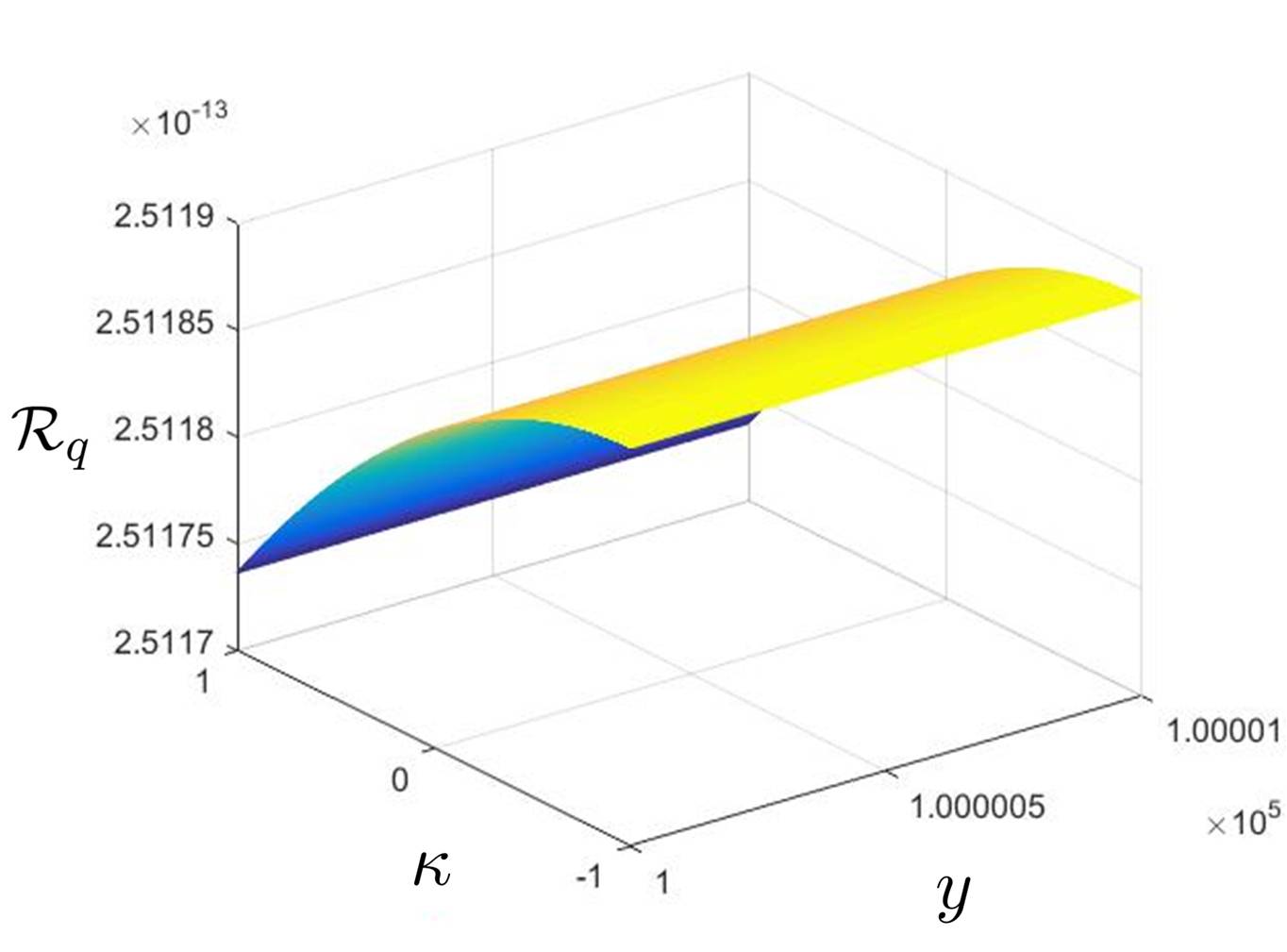}
\caption{Three-dimensional surface plot for the comoving curvature perturbation $ \mathcal{R}_q $ in a universe constructed from dust and radiation vs. rescaled sectional curvature $ \kappa $ and normalized scale factor $ y $ for the comoving wave number $ q=5 $ i.e. super-horizon scales (left) and $ q=10^{+5} $ namely severe sub-horizon modes (right) providing the adiabatic initial condition. We suppose $ n_{s_0}=0.96 $ and $ N $, the amplitude of $ \mathcal{R}_q $ at the end of inflation is roughly $ 10^{-5} $. It seems that the general behavior of $  \mathcal{R}_q $ is independent of $ q $. Notice that both $q$ and $ \mathcal{R}_q $ are dimensionless.}\label{fig2}
\end{figure}
\begin{figure}[!htb]
\centering
\includegraphics[width=.49\textwidth]{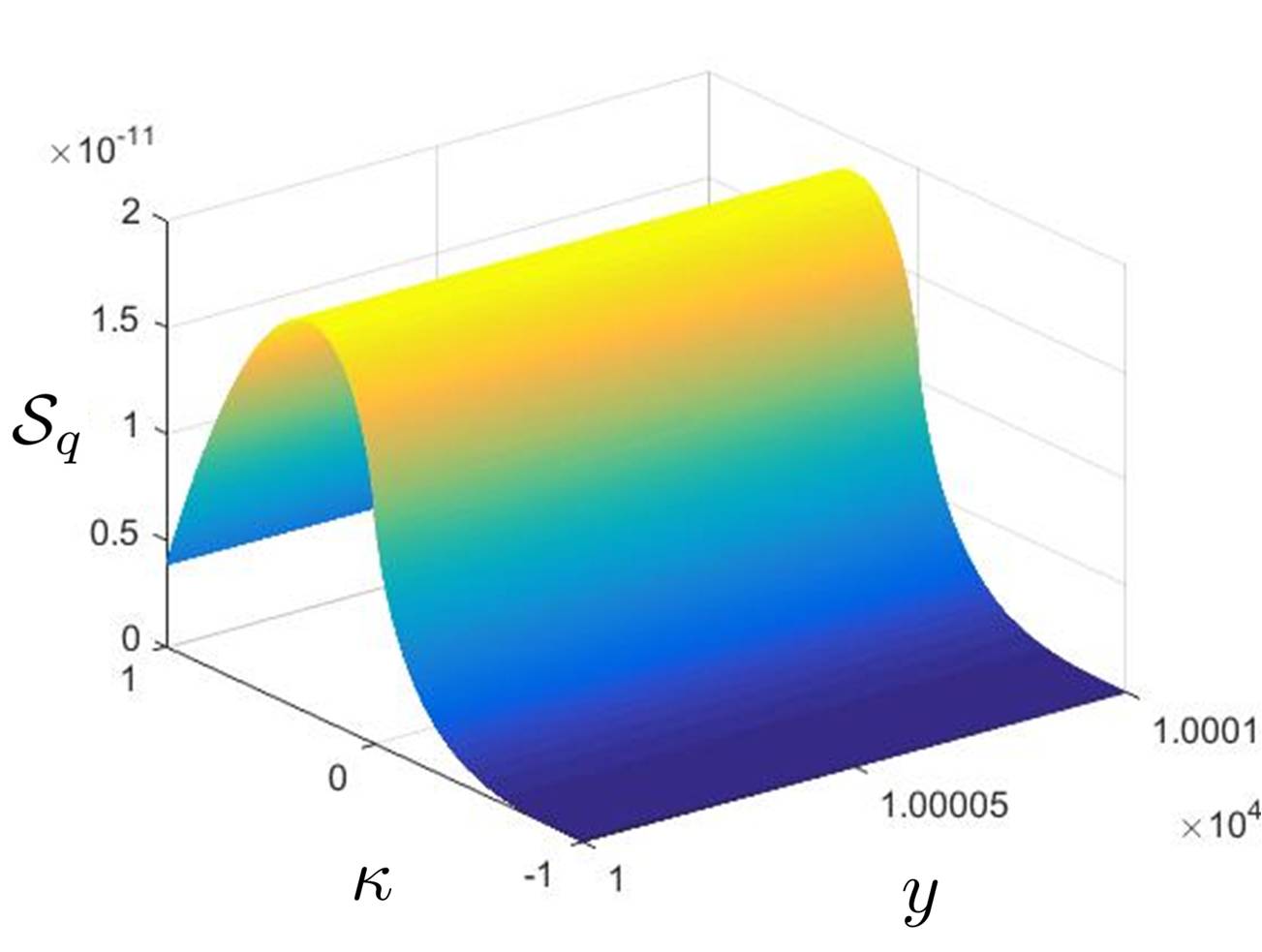}
\includegraphics[width=.49\textwidth]{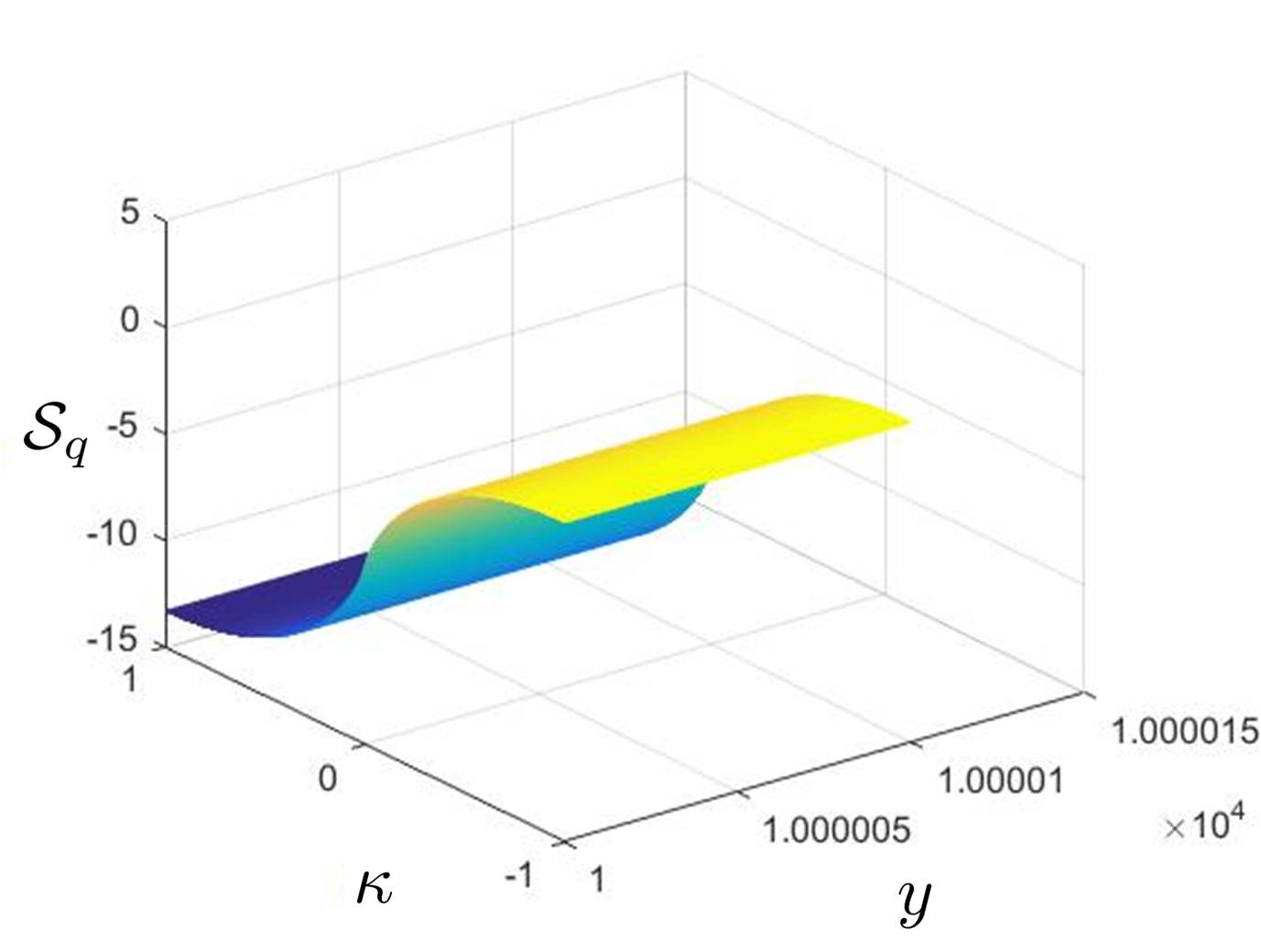}
\caption{Three-dimensional surface plot for the entropy perturbation $ \mathcal{S}_q $ in a universe constructed from dust and radiation vs. rescaled sectional curvature $ \kappa $ and  normalized scale factor $ y $ for the comoving wave number $ q=10^{+5} $ subject to the adiabatic initial condition (left) and isocurvature initial condition (right). We supposed the amplitude of $ \mathcal{S}_q $ at the end of inflation is about $ 10^{-5} $ and $ n_{iso_0}=4.43 $.}\label{fig3}
\end{figure}
The system under consideration may be solved via numerical methods like as the Runge-Kutta 4th order. The results are represented in figures \ref{fig2} and \ref{fig3} . Note that all equations have been written in terms of $ \kappa=a_{eq}^2\mathcal{K}_s $ which is \textit{rescaled sectional curvature of spatial slices of the universe}. $ \kappa $ against $ K $ is a continuous variable and can be interpreted as a topological index carries general geometrical properties, namely if $ \kappa>0 , \mathcal{M} $ topologically is equivalent to $ \mathbb{S}^3 $ otherwise $ \mathcal{M} $ is $ \mathbb{R}^3 $.\\
The spectral indices of the adiabatic and isocurvature perturbations may be defined respectively as
\begin{eqnarray}\label{a23}
&&n_s\left(q, \kappa , t\right)=4+\frac{q}{\mathcal{P}_\mathcal{R}}\frac{\partial \mathcal{P}_\mathcal{R}}{\partial q}=4+2\frac{q}{\mathcal{R}_q}\frac{\partial\mathcal{R}_q}{\partial q} ,\\
&&n_{iso}\left(q, \kappa , t\right) =4+\frac{q}{\mathcal{P}_\mathcal{S}}\frac{\partial \mathcal{P}_\mathcal{S}}{\partial q}=4+2\frac{q}{\mathcal{S}_q}\frac{\partial\mathcal{S}_q}{\partial q},
\end{eqnarray}
\begin{figure}[!htb]
\centering
\includegraphics[width=.49\textwidth]{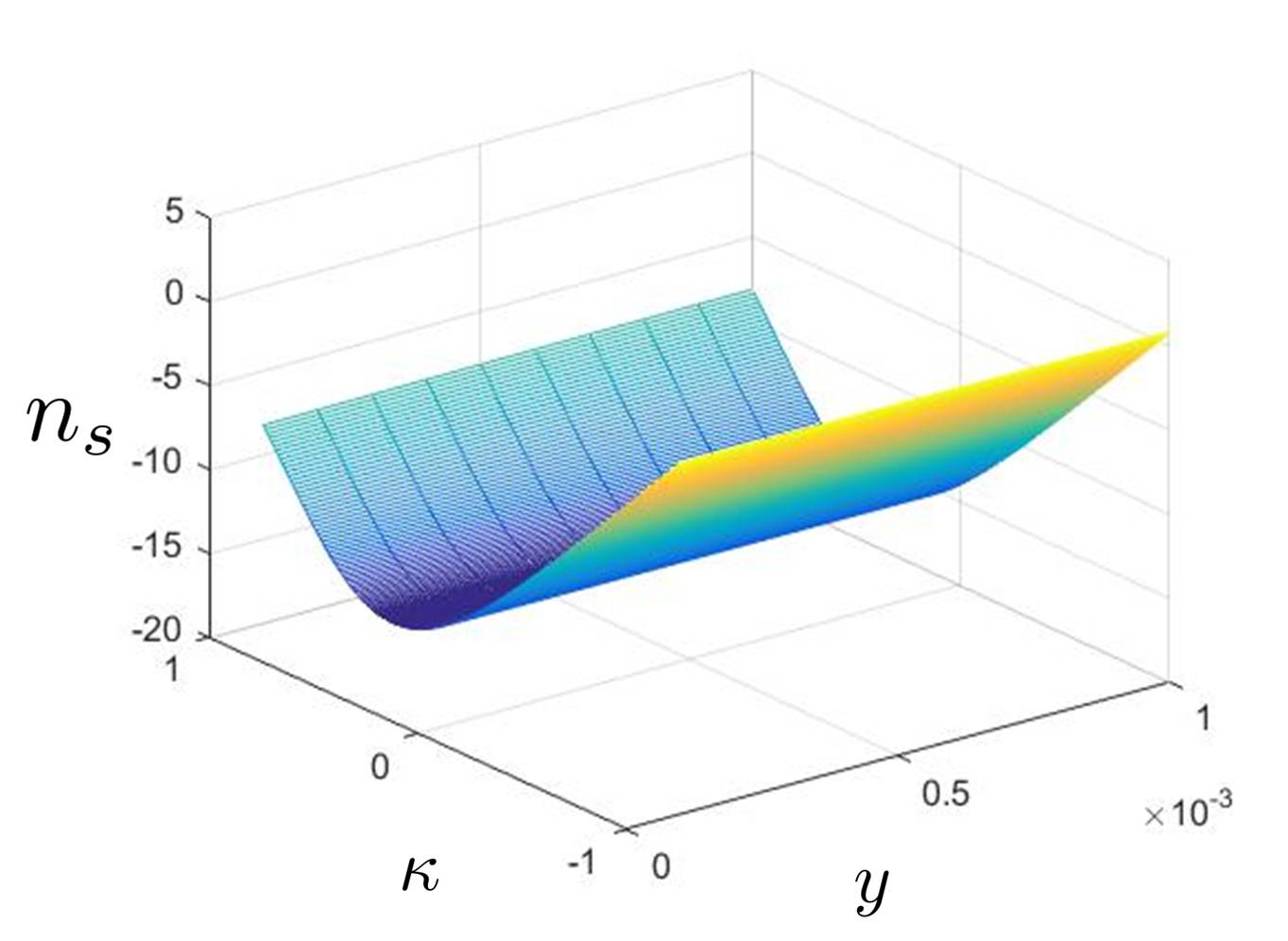}
\includegraphics[width=.49\textwidth]{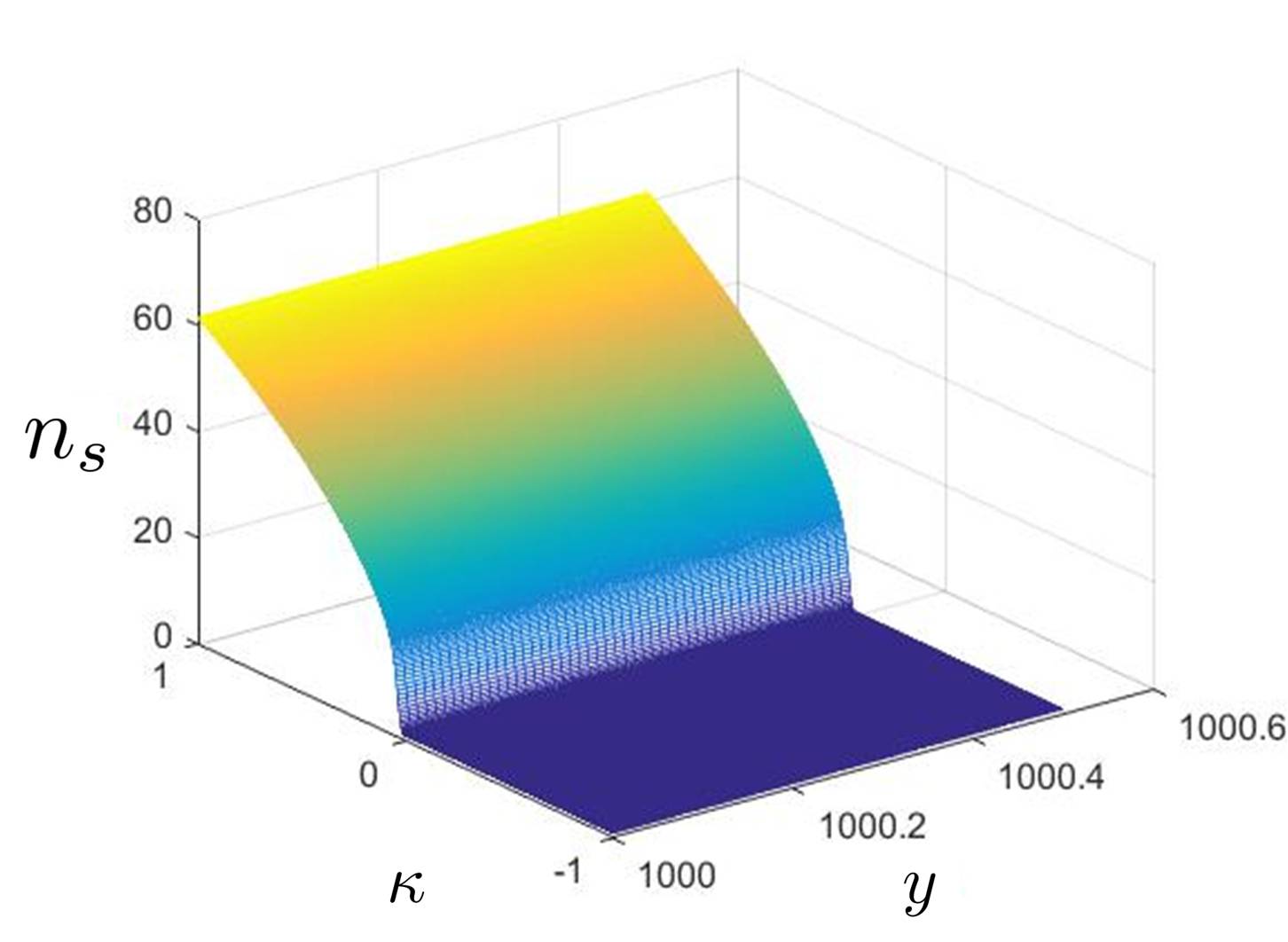}
\caption{Three-dimensional surface plot for the curvature spectral index $ n_s $ in a universe constructed from dust and radiation vs. rescaled sectional curvature $ \kappa $ and  normalized scale factor $ y $ for the comoving wave number $ q=5 $ (left) and $ q=10^{+5} $ (right) subject to the adiabatic initial condition.}\label{fig4}
\end{figure}
\begin{figure}[!htb]
\centering
\includegraphics[width=.49\textwidth]{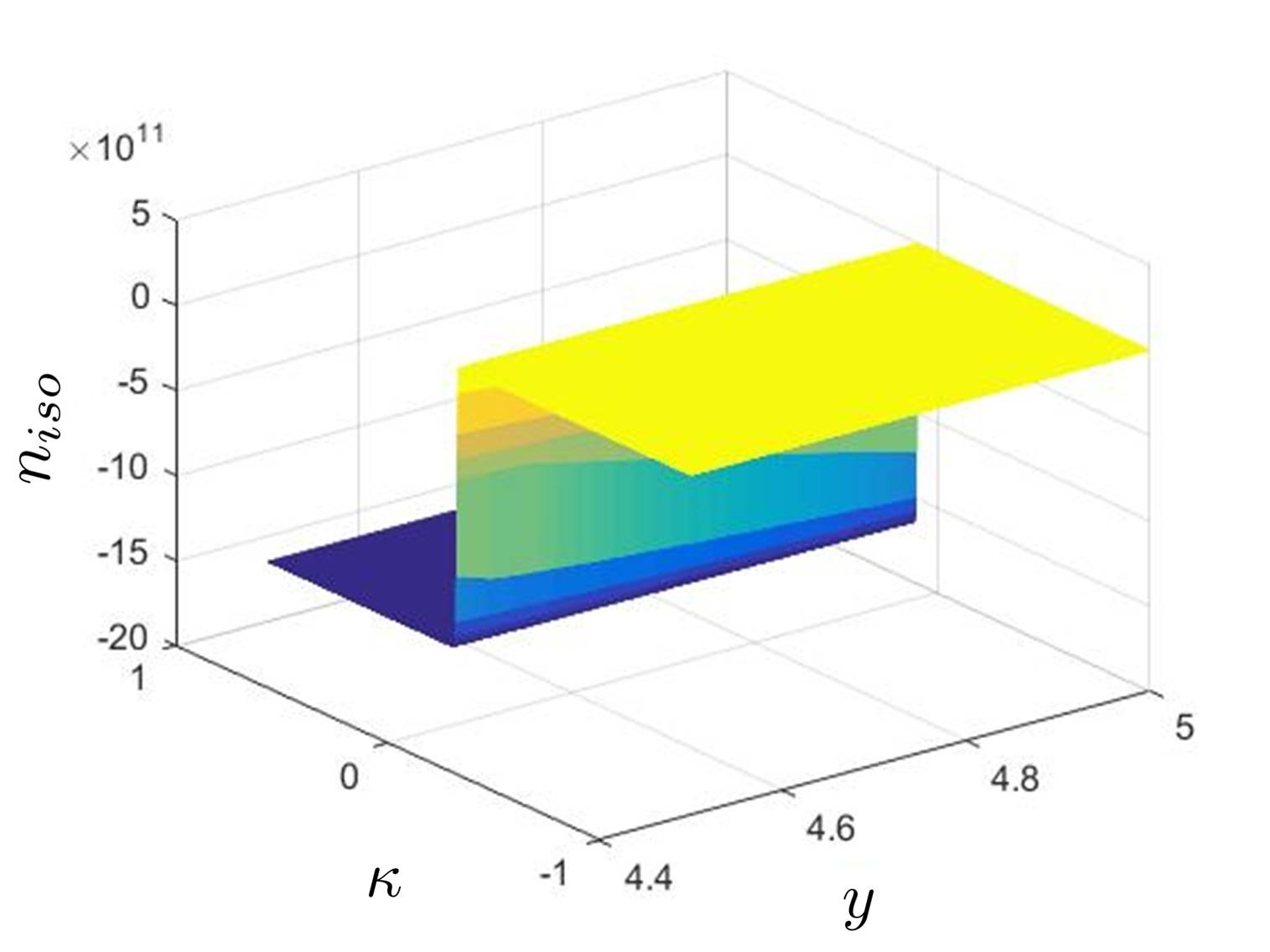}
\includegraphics[width=.49\textwidth]{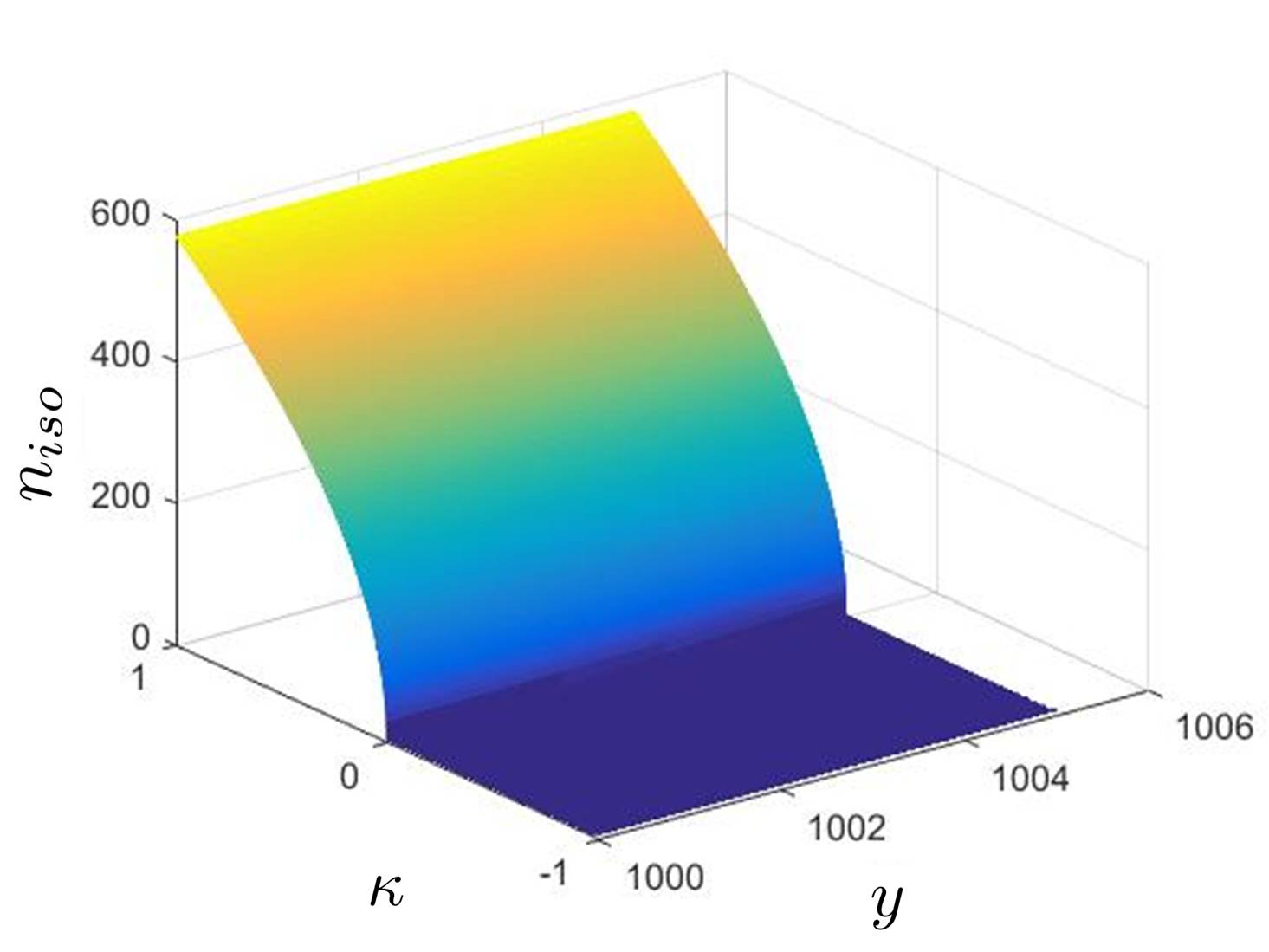}
\caption{Three-dimensional surface plot for the isocurvature spectral index $ n_{iso} $ in a universe constructed from dust and radiation vs. rescaled sectional curvature $ \kappa $ and  normalized scale factor $ y $ for the comoving wave number $ q=5 $ (left) and $ q=10^{+5} $ (right) subject to the adiabatic initial condition.}\label{fig5}
\end{figure}
\begin{figure}[!htb]
\centering
\includegraphics[width=.49\textwidth]{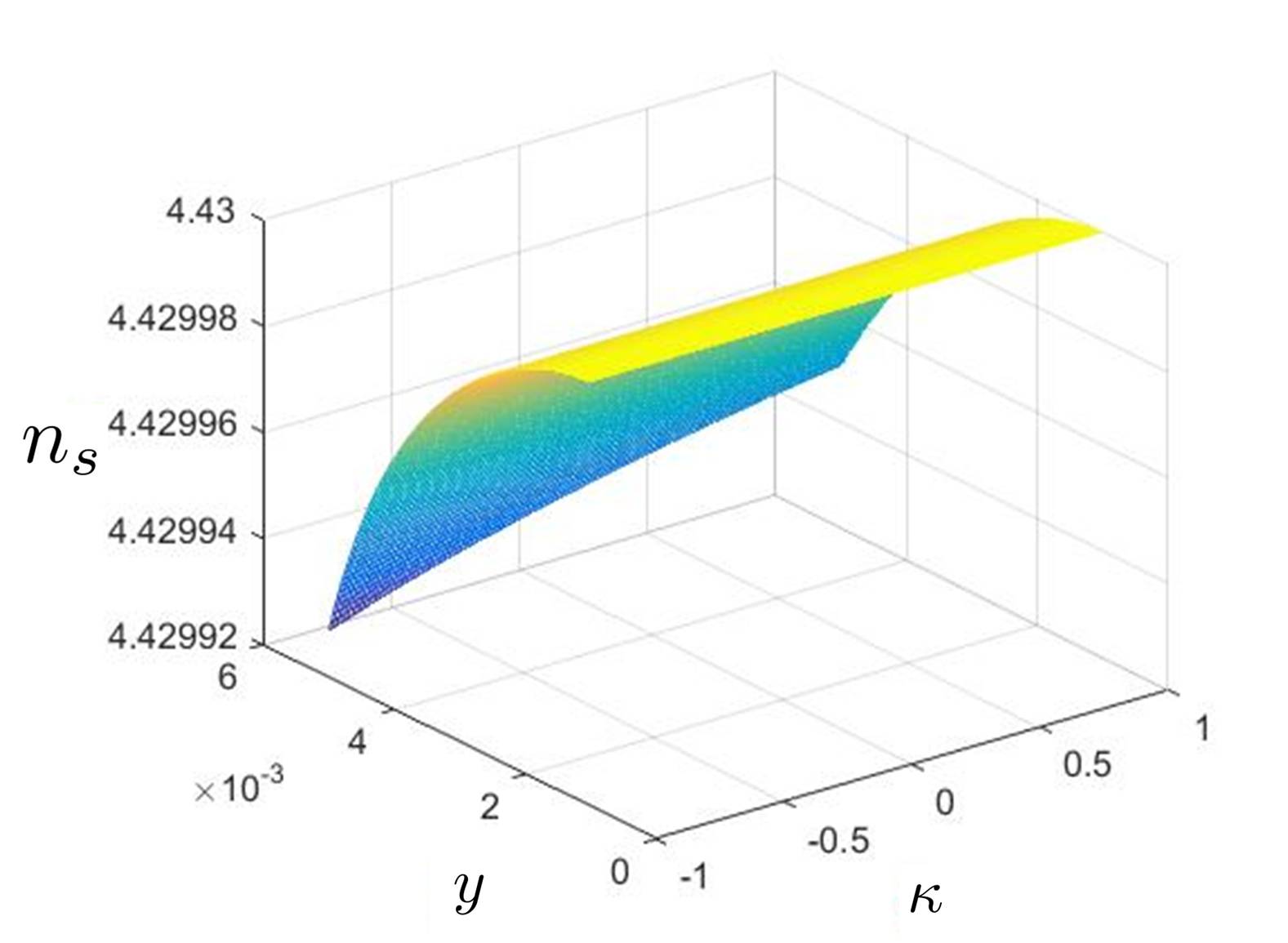}
\includegraphics[width=.49\textwidth]{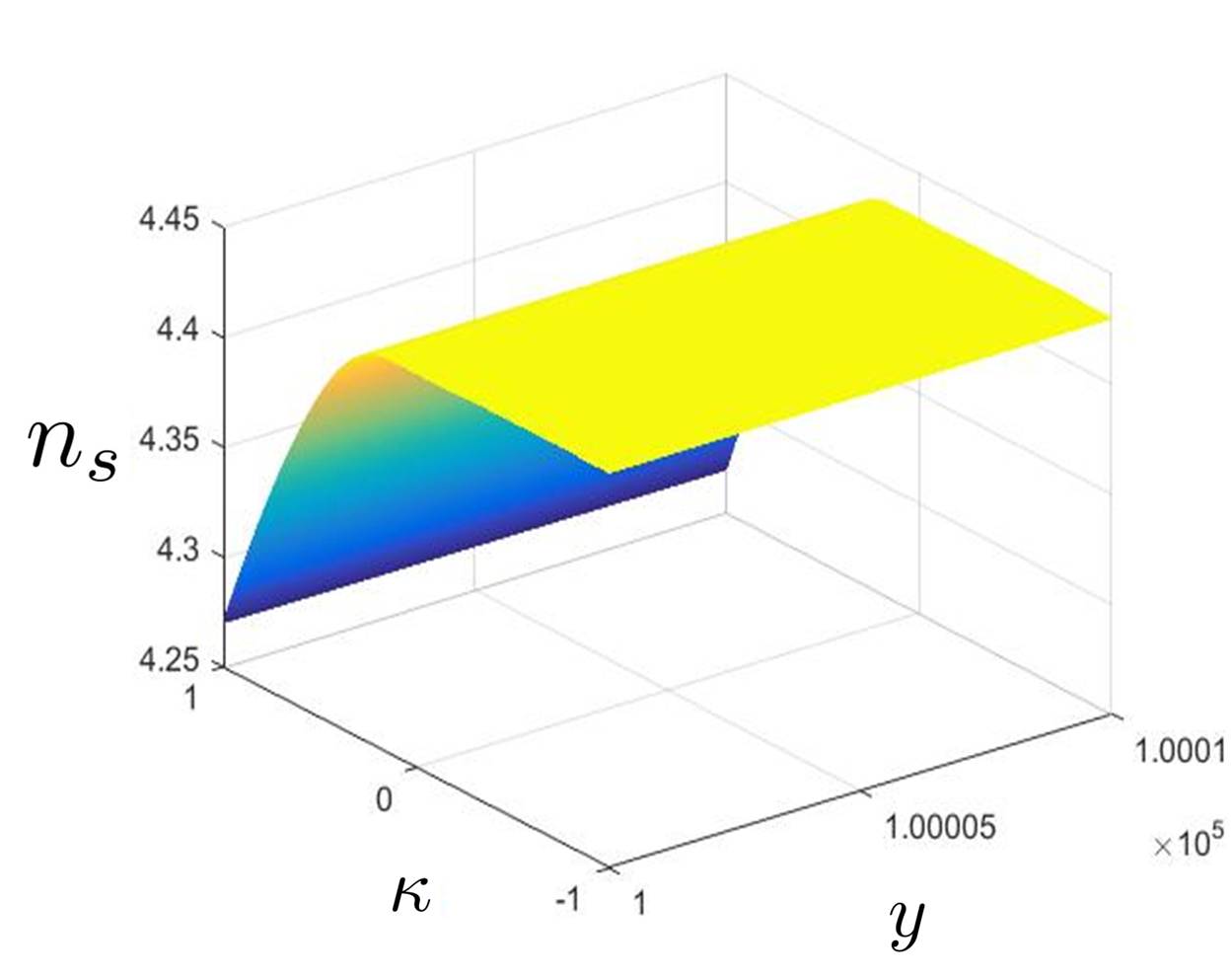}
\caption{The same as Figure \ref{fig4}, except the initial condition has changed to the isocurvature ones.}\label{fig6}
\end{figure}
\begin{figure}[!htb]
\centering
\includegraphics[width=.5\textwidth]{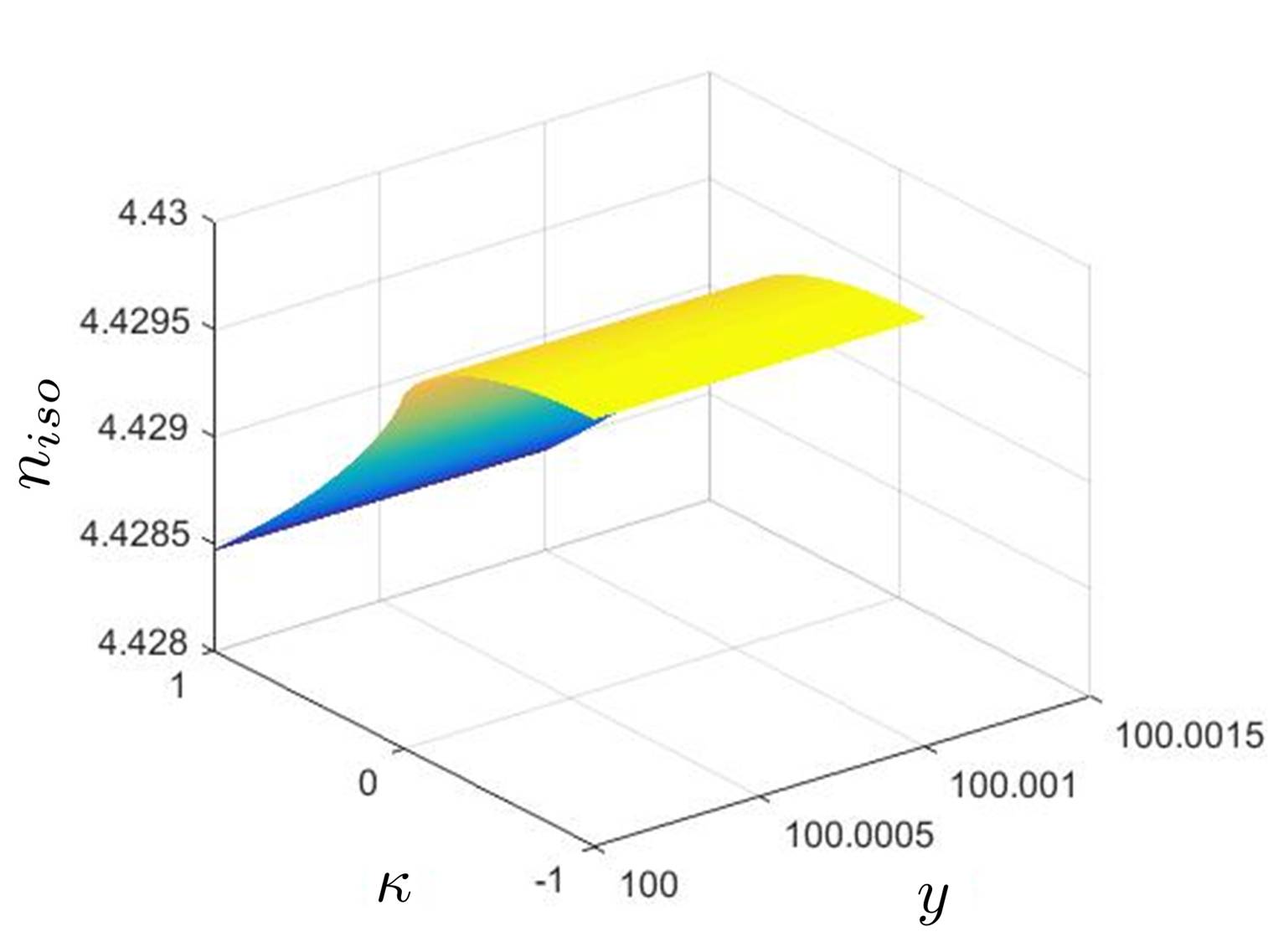}
\caption{Three-dimensional surface plot for the curvature spectral index $ n_{iso} $ in a universe constructed from dust and radiation vs. rescaled sectional curvature $ \kappa $ and  normalized scale factor $ y $ for the comoving wave number $ q=5 $ subject to the isocurvature initial condition.}\label{fig7}
\end{figure}
where $ \mathcal{P}_\mathcal{R} $ and $ \mathcal{P}_\mathcal{S} $ are the spectra of $\mathcal{R}  $ and $ \mathcal{S} $ respectively. The diagrams of $ n_s $ and $ n_{iso} $ have been shown in figures \ref{fig4} ,\ref{fig5},\ref{fig6}, and  \ref{fig7} for adiabatic and isocurvature initial conditions. It is clear that $ n_s $ and $ n_{iso} $ depend severely on $ q $, so we may say that the spectral indices are running. 
From figures \ref{fig4} and \ref{fig5}, it seems that $ n_s $ and $ n_{iso} $ for sub-horizon modes subject to adiabatic initial condition are sensitive to topology. It may be true for $  n_{iso} $ at super-horizon modes under entropic initial condition too. On the other hand, $ \mathcal{R}_q $ is inversely proportional to absolute magnitude of the sectional curvature under adiabatic initial condition. It is also true for $ \mathcal{S}_q $ at sub-horizon modes although the concavity of $ \mathcal{S}_q $-graph changes at $ \kappa =0 $ under isocurvature initial condition. Besides, at super-horizon modes subject to the adiabatic initial condition $ n_s $ is minimized at $  \kappa =0 $ nevertheless $ n_{iso} $ has inflection point. Comparison of the same figure subject to different initial conditions can illustrate the significance of the early stage in the evolution of the universe.\\
Topology-dependency of $ \mathcal{R}_q  $ as well as $ n_s $ hinges on choice of initial condition and scale of perturbations. Figures 3 and 4 make clear that $ \mathcal{R}_q$ and $ n_s $ are not sensitive to the spatial topology of the spacetime for super-horizon modes. Consequently, CMB power spectrum could not have depended on spacetime topology, because dominant perturbations at time of last scattering which free streaming of photons commenced had been outside the horizon. In other words, CMB power spectrum in a simply-connected universe cannot be affected by changing of topology. Meanwhile figure 3 indicates $ \mathcal{R}_q^o $ (the $ o $ superscript denoting ''super-horizon modes'') decreases by increasing of the spatial curvature of the universe, so the fractional density perturbations of dark matter and photons in spatially closed universe is smaller than flat case\cite{6} although structure can be formed more rapidly as mentioned before. In another front, Figures 5-7 clearly illustrate the changes of spectral indices over time is permissibly negligible and in accordance with observational data \cite{11} the values $ n_{s0} $ and $ n_{iso0} $ may be admitted for all times. So by an accurate probing figure 7 (right) and comparing to the released observational data about value of $ n_s $ \cite{11} one can deduce spatial curvature of universe must have been $ -0.0190\lesssim\kappa\lesssim+0.005 $.   

\section{Super-curvature modes}
\label{sec5}

\begin{figure}[!htb]
\centering
\includegraphics[width=.49\textwidth]{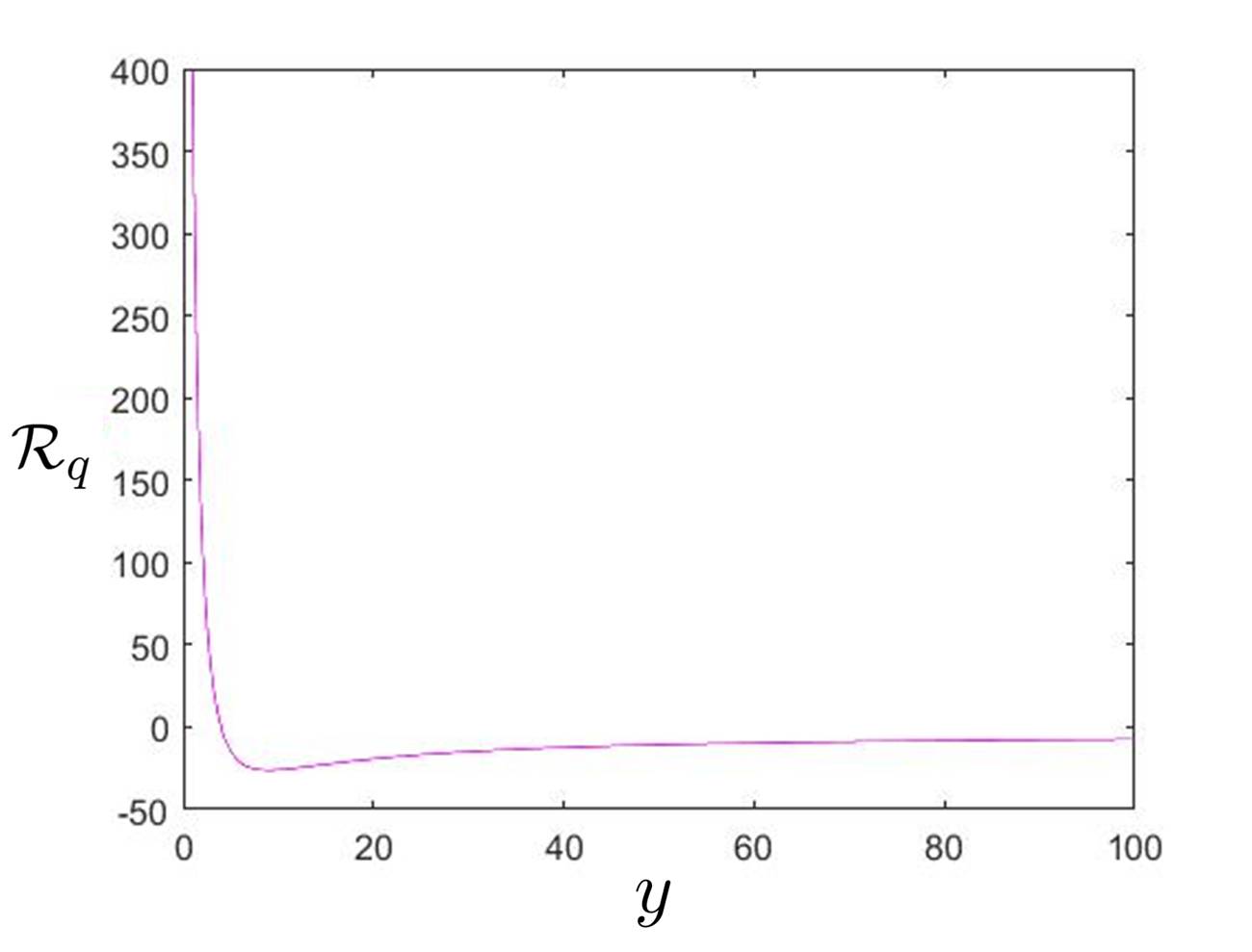}
\includegraphics[width=.49\textwidth]{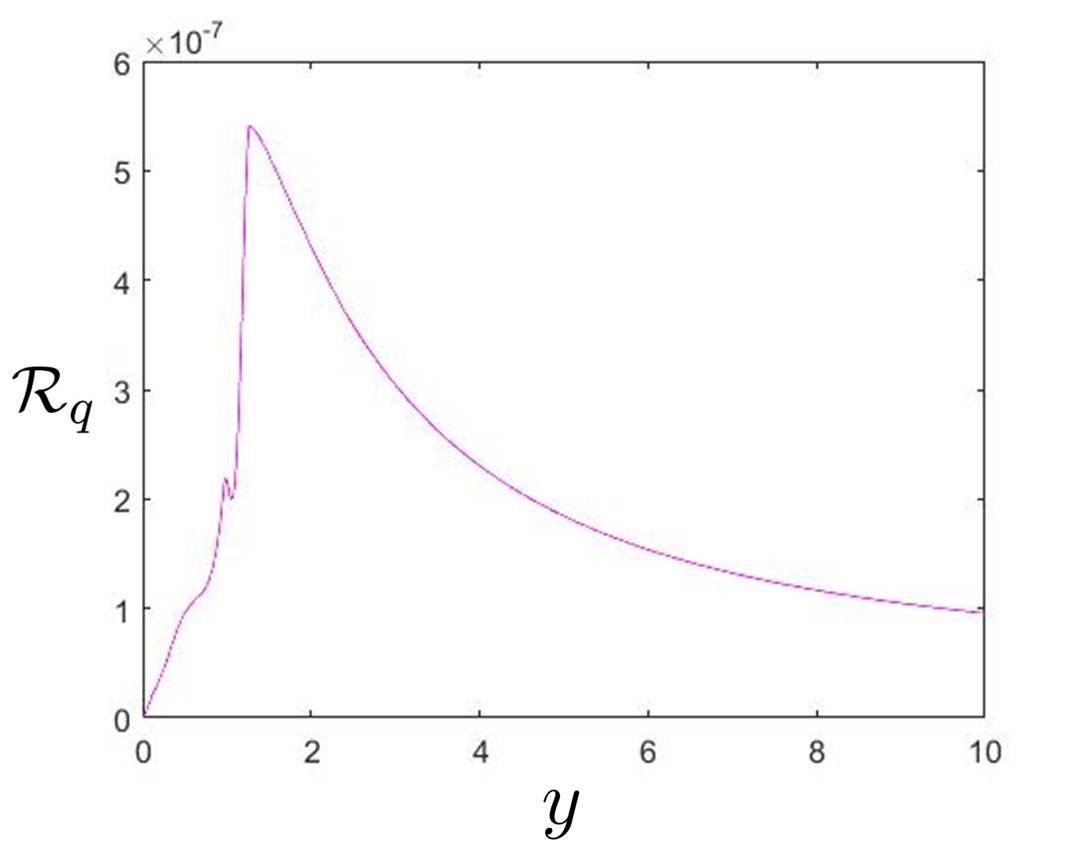}
\caption{Evolution of the comoving curvature perturbation $ \mathcal{R}_q $ in a universe constructed from dust and radiation for a typical super-curvature scale subject to adiabatic (left) and isocurvature (right) initial conditions.}\label{fig8}
\end{figure}

\begin{figure}[!htb]
\centering
\includegraphics[width=.49\textwidth]{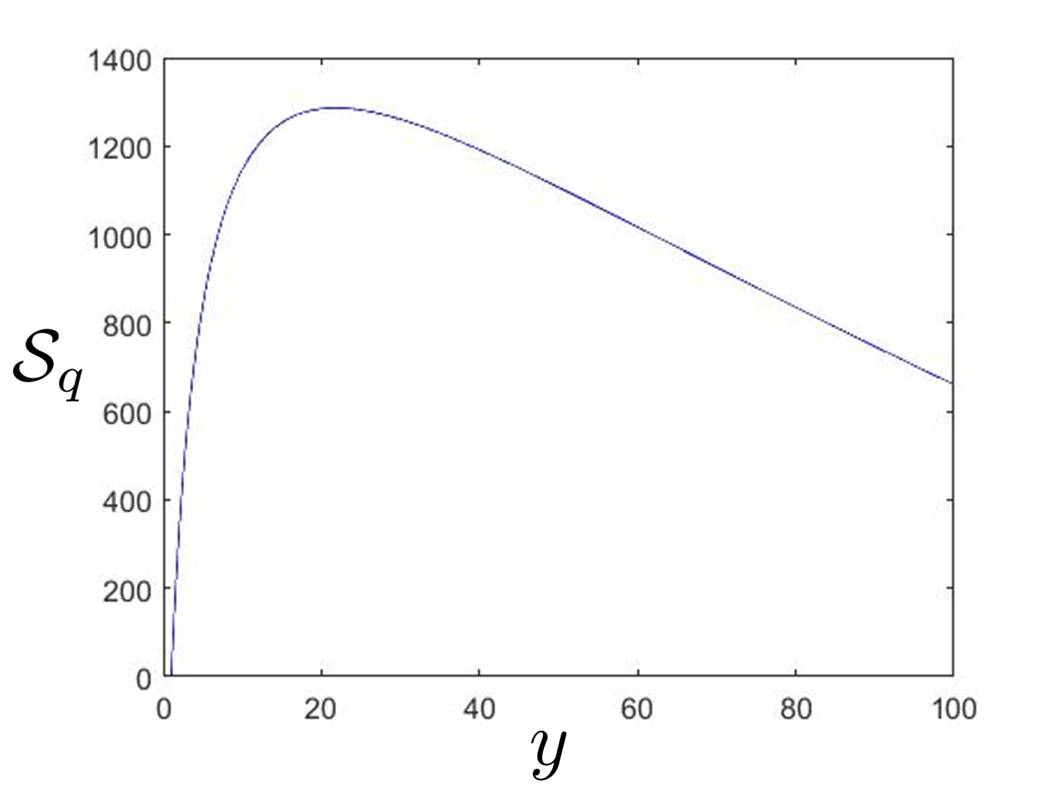}
\includegraphics[width=.49\textwidth]{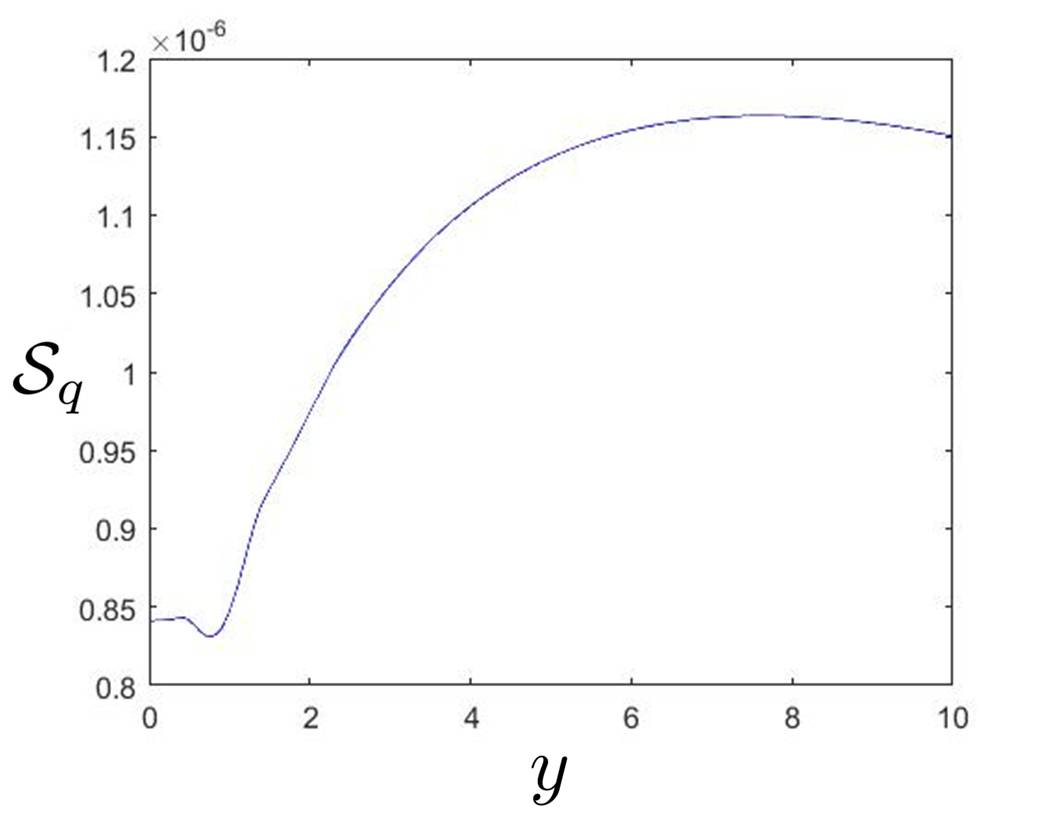}
\caption{The same as Figure \ref{fig8}, exept for the entropy perturbation $ \mathcal{S}_q $.}\label{fig9}
\end{figure}

\begin{figure}[!htb]
\centering
\includegraphics[width=.49\textwidth]{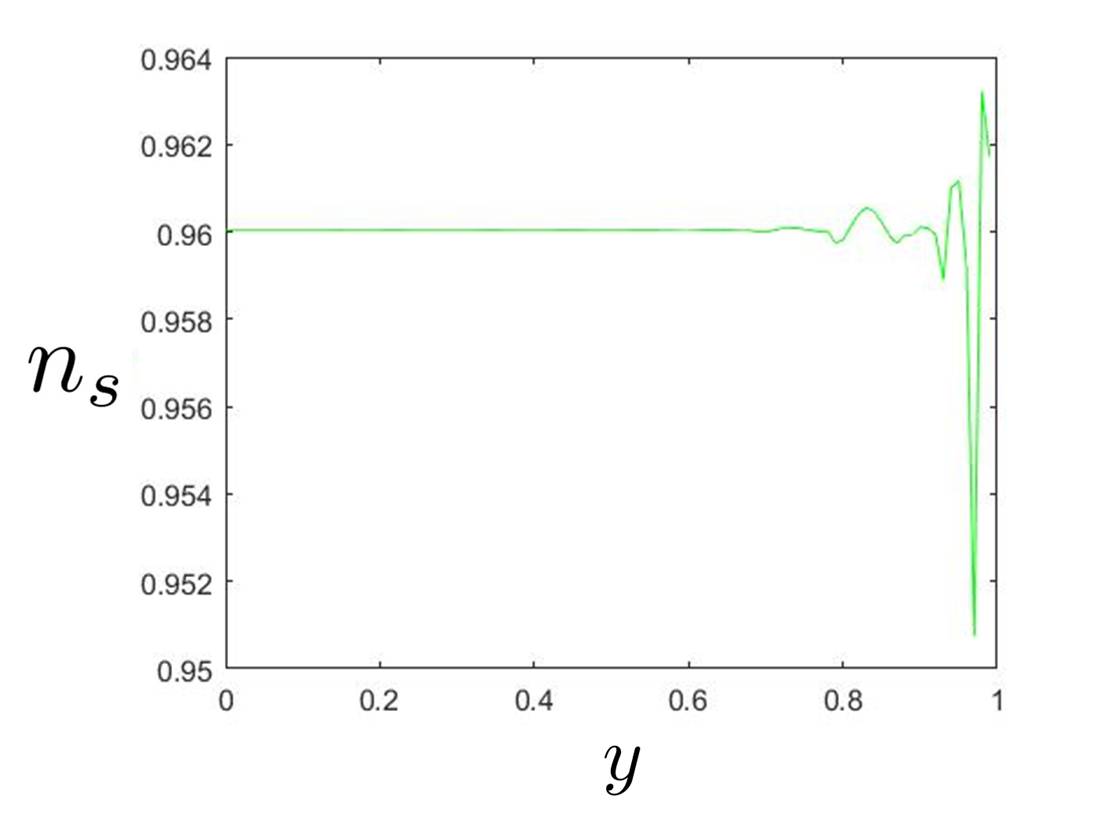}
\includegraphics[width=.49\textwidth]{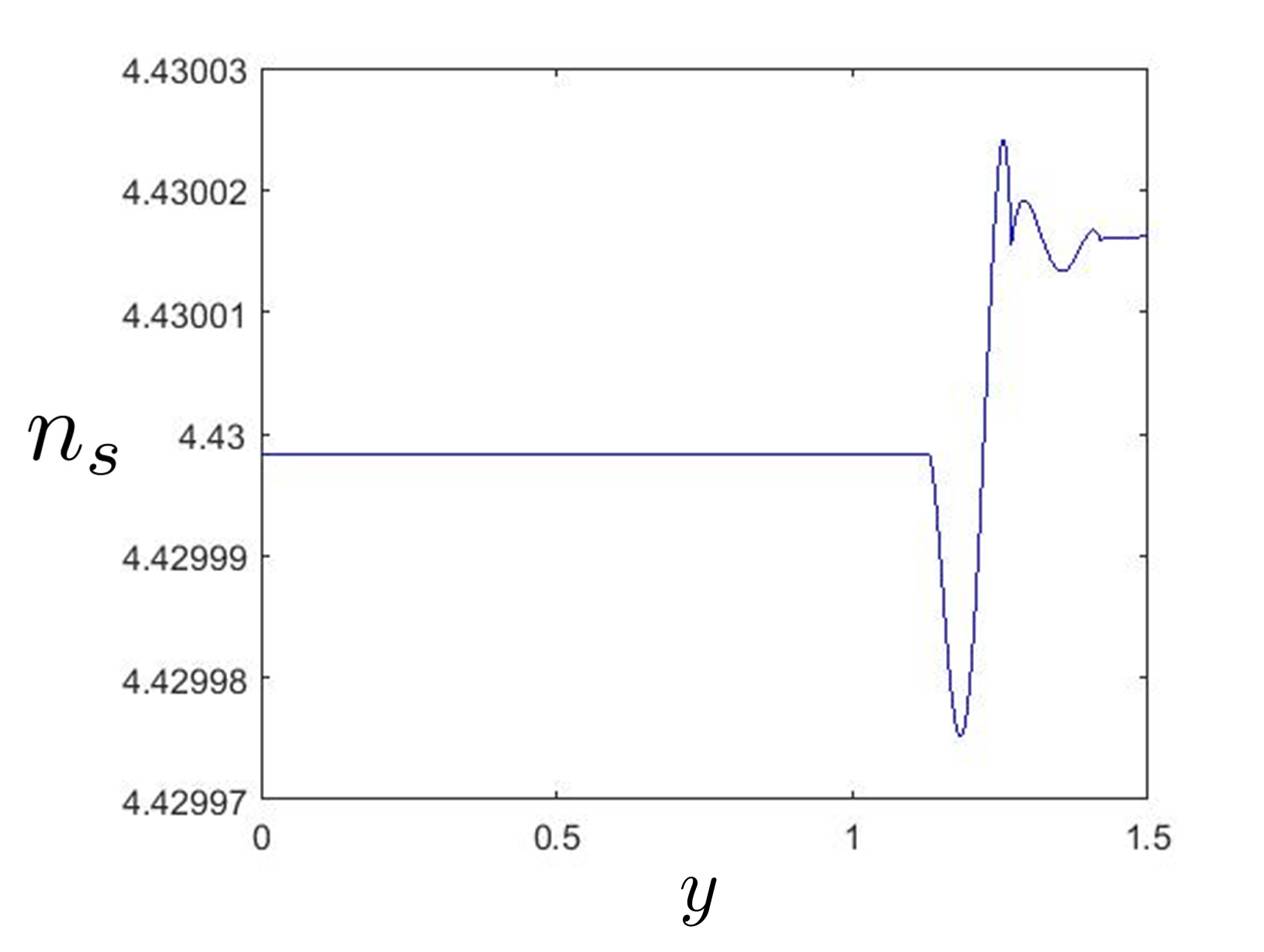}
\caption{The same as Figure \ref{fig8}, exept for the curvature spectral index $ n_s $.}\label{fig10}
\end{figure}

\begin{figure}[!htb]
\centering
\includegraphics[width=.49\textwidth]{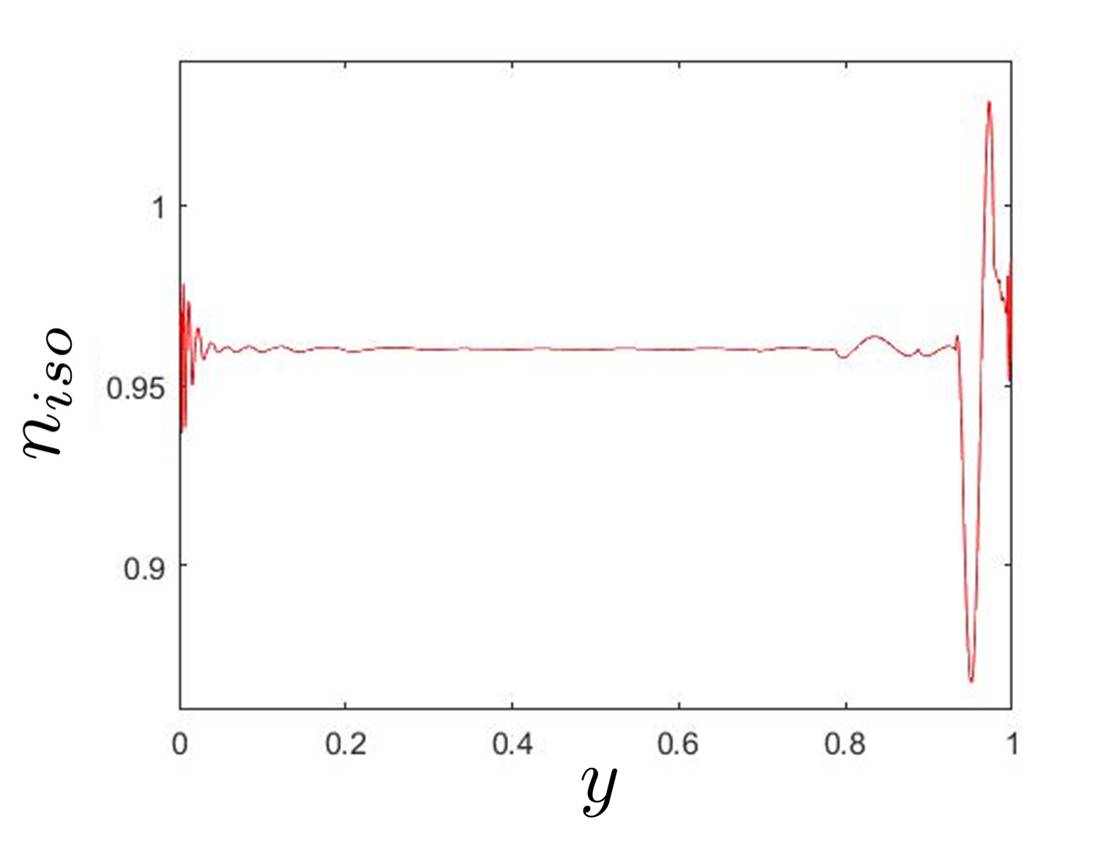}
\includegraphics[width=.49\textwidth]{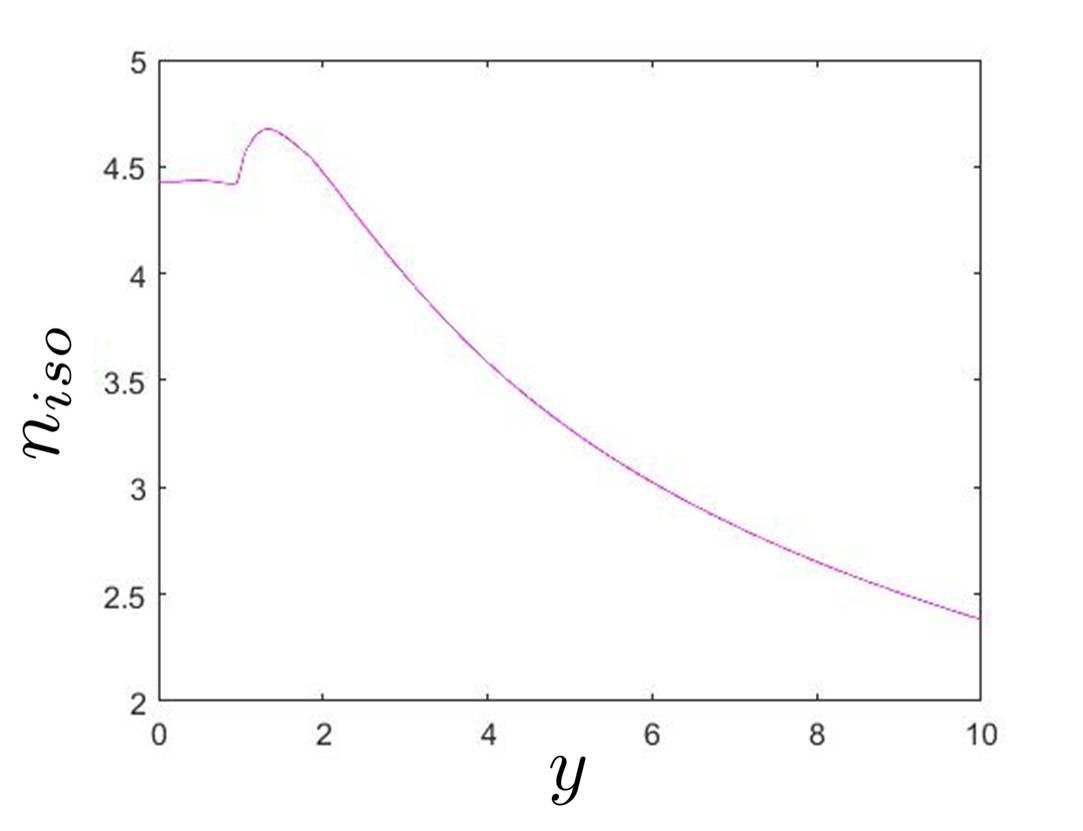}
\caption{The same as Figure \ref{fig8}, exept for the isocurvature spectral index $ n_{iso} $.}\label{fig11}
\end{figure}

In case $ K=-1 $ (open universe) eigenvalues of the Laplace-Beltrami operator are
\begin{equation}\label{a24}
\mathfrak{K}_q=-1-q^2
\end{equation}
We may suppose $ 0<q^2<+\infty $ which results in sub-curvature modes. On the contrary, someone may suppose $ -1<q^2<0 $ from which super-curvature modes come out. Super-curvature modes appear merely in the open universe and may be the solutions of odd phenomena observed in CMB anisotropy\cite{23}. So we investigate evolution of $ \mathcal{R}_q $, $ \mathcal{S}_q $, $ n_s $, and $ n_{iso} $ for the super-curvature modes under different initial conditions separately. The results can be observed in figures \ref{fig8}-\ref{fig11}.
\section{Discussion and Conclusion}
\label{sec6}

We have derived a neat equation (generalized Mukhanov-Sasaki equation) for the evolution of comoving curvature perturbation as a leading random field in the cosmological perturbation Theory. This equation obviously divulges dependency of the comoving curvature perturbation to the curvature index of the FLRW universe. We also considered the numerical solutions of this equation for the mixture of dust and radiation subject to adiabatic and isocurvature initial conditions. As we have seen, in question equation cannot be solved alone due to the presence of another random field i.e. entropy perturbation, so we are coerced to solve the Mukhanov-Sasaki equation along with the Kodama-Sasaki equation simultaneously. We investigated the time evolution of the adiabatic and entropy spectral indices under different initial conditions too. It sounds that both spectral indices for perturbations with scales deep inside the horizon under adiabatic initial condition are severely sensitive to topology-changing. On the other hand, for the super-horizon modes this behavior is not observed and consequently CMB power spectrum in the simply-connected universe must be independent of topology. Moreover, we found that $ n_s $ decreases when the universe becomes more flat. It is also clear that in case $ K=0 $, $n_s $ has an increasing rate for super-horizon modes in early times regardless of the initial conditions. Besides, the oscillating behavior of cosmological perturbations and their indices in case $K=+1 $ for all scales is thoroughly clear. We also concluded cosmic structure in the spatially closed case must have been formed more rapidly. Finally, we examined evolution of $ \mathcal{R}_q $, $ \mathcal{S}_q $ and their spectra for the super-curvature perturbation for which $ K=-1 $ too. 
\appendix
\section{Derivation of the generalized Mukhanov-Sasaki equation}
\label{sec:6}
The geometry of the universe may be described by
\begin{equation}\label{a25}
g_{\mu \nu }  = \bar{g}_{\mu \nu }  + h_{\mu \nu } ,
\end{equation}
$ \bar{g}_{\mu \nu } $ Being the unperturbed FLRW metric\cite{7}
\begin{equation}\label{a26}
 g_{00}  =  - 1 ,\quad g_{0i}  = g_{i0}  = 0 ,\quad g_{ij}  = a^2 \left( t \right)\tilde g_{ij}=a^2 \left( t \right)\left( \delta _{ij}  + K\frac{x^i x^j}{1 - K{\bf{x}}^2}\right) ,
\end{equation}
and $  h_{\mu \nu } $ is a small perturbation can be parameterized as\cite{7}
\begin{equation}\label{a27}
h_{\mu \nu }dx^\mu dx^\nu  =-Edt^2+2a\partial_i F dt dx^i+a^2\left(A\tilde{g}_{ij}+\mathcal{H}_{ij} B\right)dx^i dx^j,
\end{equation}
where the perturbations $ A $, $ B $, $ E $ and $ F $ are time-dependent random fields (here we condone vector and tensor parts of perturbations). On the other hand, energy-momentum of the cosmic fluid is given by\cite{7}
\begin{eqnarray}
&&T_{00}= \bar \rho\left( 1 + E\right)  + \delta\rho ,\\
&&T_{i0} = a\bar p \partial_i F-\left(\bar \rho+\bar p\right)\partial _i\left(\delta u \right) ,\\
&&T_{ij} = a^2\left[\bar p\left(1+A \right)\tilde{g} _{ij}+\delta p\tilde{g}_{ij}+ \mathcal{H}_{ij} \left(\bar p B +\Pi ^S\right) \right] . \label{a2}
\end{eqnarray}
Now the Einstein field equations along with conservation law lead to a system of coupled linear differential equations (see equations 20-25 in\cite{7}) which one can rewrite them in terms of gauge-invariant random fields introduced in section 2 in order to get rid of gauge freedom:
\begin{align}
&\Psi  - \Phi  = 8\pi Ga^2\Pi ^S, \label{a28} \hfill \\
&\dot{\Psi}  + H\Phi  =  - 4\pi G\left(\bar{\rho}  + \bar{p} \right)V,  \label{a29} \hfill \\
&\nabla ^2\Phi  + 3H\dot{\Phi}+ 6\left( H^2+ \dot{H} \right)\Phi  + 3\ddot{\Psi} + 6H\dot{\Psi} = 4\pi G\left[\dot{\bar{ \rho}} \left(1 + 3{c_s}^2 \right)D_s + 3\Gamma  + a^2\Pi ^S \right],   \label{a30}\hfill \\
&-\frac{4K}{a^2}\Psi + 6H\dot{\Psi} +\ddot{\Psi} -\nabla ^2\Psi  + 2\left( 3H^2 + \dot{H} \right)\Phi  + H\dot{\Phi}  = 4\pi G\left[ \dot{\bar{\rho}}\left( {c_s}^2 - 1 \right){D_s} + \Gamma  +a^2\Pi ^S\right],   \label{a31}\hfill \\
&\left(\bar{\rho}+\bar{p}\right)\nabla ^2 V +a^2H\nabla ^2\Pi ^S + 3H\Gamma  + 3\left(\bar{\rho} + \bar{p}\right)\dot{\zeta} = 0,  \label{a32}\hfill \\
&\dot{\bar{\rho}} {c_s}^2\Delta + \left(\bar{\rho} +\bar{p}\right)\left(\Phi  + \dot{V} \right) + \Gamma +a^2\nabla ^2\Pi ^S + 2K\Pi ^S = 0, \label{a33}
\end{align}
where $ D_s=-\frac{1}{H}\left(\zeta+\Psi \right)  $ and $ H $ and is the Hubble parameter. By combination of equations (\ref{a28}), (\ref{a29}) and (\ref{a31}) Bardeen equation specifying dynamics of $ \Psi $ is obtained
\begin{multline}\label{a34}
\ddot{\Psi}+ H\left(4 + 3{c_s}^2\right)\dot{\Psi}+\left[3H^2\left(1 + {c_s}^2\right) + 2\dot{H}- \frac{K}{a^2}\left(1 + 3{c_s}^2\right)\right]\Psi - {c_s}^2\nabla ^2\Psi = \\
  4\pi G\left[\Gamma + a^2\nabla ^2\Pi ^S+ 2Ha^2\dot{\Pi }^S + 2a^2\left(5H^2 + 2\dot{H} + 3H^2{c_s}^2\right)\Pi ^S\right] .
\end{multline}
On the other hand, one can put $ V=\frac{\mathcal{R}+\Psi}{H} $ in equation (\ref{a29}) and use equation (\ref{a28}) to obtain
\begin{equation}\label{a35}
\mathcal{R} = \frac{\dot{H} -H^2 - \frac{K}{a^2}}{4\pi G\left(\bar{\rho}  + \bar{p}\right)}\Psi - \frac{H}{4\pi G\left(\bar {\rho}  + \bar{p}\right)}\dot{\Psi}  + \frac{2H^2a^2}{\bar{\rho}  + \bar{p}}\Pi ^S.
\end{equation}
By taking time derivation of equation (\ref{a35}) as well as using equation (\ref{a34}) and also relation
\begin{equation}
\frac{d}{dt}\left(\bar{p}+\bar{\rho} \right)=-3H \left(\bar{p}+\bar{\rho} \right)\left(1+{c_s}^2 \right),
\end{equation}
we have
\begin{equation}\label{a36}
\dot{\mathcal{R}} = \frac{H{c_s}^2}{\dot{H} - \frac{K}{a^2}}\nabla ^2\Psi  + \frac{4\pi GH\left(\Gamma  +a^2\nabla ^2{\Pi ^S} \right)}{\dot{H} - \frac{K}{a^2}} + \frac{K}{\dot{H}a^2 - K}\left[\dot{\Psi} + H\left(1 + 3{c_s}^2\right)\Psi \right] .
 \end{equation}
Now by invoking conformal time $ \tau=\int^t_{t_0}\frac{d\xi}{a\left( \xi\right) } $, equations (\ref{a35}) and (\ref{a36}) reduce to equations (\ref{a3}) and (\ref{a4}). Meanwhile by taking Fourier transform from equations (\ref{a3}) and (\ref{a4}), equations (\ref{a5}) and (\ref{a6}) could be derived. \\
By elimination once $ \Psi_q $ and again $ \Psi'_q $ between equations (\ref{a5}) and (\ref{a6}) separately one can obtain  
\begin{equation}\label{a37}
\Psi_q=-\frac{\mathscr{A}\left(K\mathcal{R}_q-\mathcal{H}\mathcal{R}'_q \right)}{\mathscr{B}_q}+\frac{4\pi G\mathcal{H}^2a^2\mathscr{C}_q}{\mathscr{B}_q}  ,
\end{equation}
and
\begin{multline}\label{a38}
\Psi'_q=-K\mathscr{A}\mathcal{R}'_q-\frac{K\mathscr{A}\mathcal{H}\left[ {c_s}^2\left(q^2-4K \right)-K \right]\left(K\mathcal{R}_q-\mathcal{H}\mathcal{R}'_q \right)}{\mathscr{B}_q}+\\
\frac{4\pi GK\mathcal{H}^3a^2\left[ {c_s}^2\left(q^2-4K \right)-K \right]\mathscr{C}_q}{\mathscr{B}_q}-4\pi G K\mathcal{H}a^2\left[\Gamma_q-\left(q^2-K \right)\Pi^S_q \right] , 
\end{multline}
where $ \mathscr{A} $, $ \mathscr{B}_q $ and $ \mathscr{C}_q $ are defined in section 2. Now by insertion $ \Psi_q $ from equation (\ref{a37}) in the left-hand side of equation (\ref{a38}), equation (\ref{a7}) can be deduced.


\bibliographystyle{elsarticle-num}
\bibliography{<your-bib-database>}



\end{document}